\documentclass[12pt]{article}
\pdfoutput=1       
 
\usepackage{hyperref}     
\usepackage{amsmath}        
\usepackage{amssymb} 
\usepackage{graphicx}
\usepackage{url}
\usepackage{enumerate}
\usepackage{color}
\usepackage{ulem}
\usepackage{tocloft}
\usepackage{float,wrapfig,color} 

\allowdisplaybreaks[1]
\def\eq#1{(\ref{#1})}
\def\s[#1\s]{\begin{align}\begin{split}#1\end{split}\end{align}}
\def\[#1\]{\begin{align}#1\end{align}}

\def\bpsi{{\bar \psi}}
\def\bvphi{{\bar \varphi}}
\def\vphi{{\varphi}}

\def \bpsit{{\bar \psi^\perp}}
\def\psit{\psi^\perp}

\def\tb{\tilde b}
\def\ta{\tilde a}
\def\tc{\tilde a^*}
\def\as{a^*}

\def\vr{v_R}
\def\vi{v_I}
\def\tvr{\tilde v_R}
\def\tvi{\tilde v_I}
\def\tv{\tilde v}
\def\av{|v|}

\def\re#1{{\rm Re}(#1)}
\def\im#1{{\rm Im}(#1)}

\def\dperp{\perp \mkern-10mu \perp}

\begin{document}

\begin{titlepage} 

\title{
\hfill\parbox{4cm}{ \normalsize YITP-24-83}\\   
\vspace{1cm} 
Three cases of complex eigenvalue/vector distributions \\
of symmetric order-three random tensors}

\author{
Swastik Majumder$^1$\footnote{sm20ms175@iiserkol.ac.in, majumderswastik09@gmail.com}
and 
Naoki Sasakura$^{2,3}$\footnote{sasakura@yukawa.kyoto-u.ac.jp}
\\
$^1${\small{\it Department of Physical Sciences, Indian Institute of Science Education and Research Kolkata, }}\\
{\small{\it Campus Rd, Mohanpur, Haringhata Farm, West Bengal 741246, India}}\\
$^2${\small{\it Yukawa Institute for Theoretical Physics, Kyoto University, }}\\
{\small{\it Kitashirakawa, Sakyo-ku, Kyoto 606-8502, Japan}}\\
$^3${\small{\it CGPQI, Yukawa Institute for Theoretical Physics, Kyoto University,}} \\
{\small{\it Kitashirakawa, Sakyo-ku, Kyoto 606-8502, Japan}}
}
  
\date{\today} 
 
 
\maketitle 
\begin{abstract}
Random tensor models have applications in a variety of fields, such as quantum gravity, quantum information theory, 
mathematics of modern technologies, etc., 
and studying their statistical properties, e.g., tensor eigenvalue/vector distributions, 
are interesting and useful. 
Recently some tensor eigenvalue/vector distributions have been computed by expressing
them as partition functions of zero-dimensional quantum field theories. 
In this paper, using the method, 
we compute three cases of complex eigenvalue/vector distributions of symmetric order-three 
random tensors, where the three cases can be characterized by the Lie-group invariances,
$O(N,\mathbb{R})$, $O(N,\mathbb{C})$, and $U(N,\mathbb{C})$, respectively.
Exact closed-form expressions of the distributions are obtained by computing partition functions of four-fermi theories, 
where the last case is of the ``signed" 
distribution which counts the distribution with a sign factor coming from a Hessian matrix. 
As an application, we compute the injective norm of the complex symmetric order-three random tensor in the large-$N$ limit
by computing the edge of the last signed distribution, obtaining agreement with a former numerical result 
in the literature. 

\end{abstract}

\thispagestyle{empty}

\end{titlepage}  

\tableofcontents

\newpage

\section{Introduction}
\label{sec:introduction}
 
Eigenvalue distributions are important quantities in the applications of random matrix models. Wigner modeled
Hamiltonians of nuclei as random matrices and obtained the celebrated semi-circle law of the eigenvalue 
distribution \cite{Wigner}.
Eigenvalue distributions play vital roles in solving random matrix models \cite{Brezin:1977sv,matrix}. 
Topological transitions of eigenvalue distributions provide stimulating insights into the QCD dynamics 
\cite{Gross:1980he,Wadia:1980cp}.     
 
The notion of eigenvalue/vector can be extended to tensors \cite{Qi,lim,cart,qibook}. 
They are defined by a coefficient and a vector (or vectors)
which satisfy an equation (or a system of equations) similar to an eigenvalue/vector equation of a matrix. 
It is non-linear with respect to an eigenvector(s), unlike in the matrix case, 
and appears in a variety of contexts, 
such as quantum \cite{Biggs:2023mfn}/classical \cite{Evnin:2021buq} gravity, 
spin glasses \cite{pspin,pedestrians,randommat}, 
computer sciences \cite{SAPM:SAPM192761164,Carroll1970,comon,spiked,bestrankone}, 
quantum information \cite{shi,barnum,geommeasure,estimate} and more \cite{qibook}, 
even though the terminology, tensor eigenvalue/vector, is not necessarily used in these contexts. 

While a matrix eigenvalue/vector can systematically be computed by the standard methods, 
computing a tensor eigenvalue/vector (or vectors) is known to be NP-hard \cite{nphard}. 
On the other hand, statistical properties of tensor eigenvalues/vectors, such as distributions of eigenvalues/vectors 
of random tensors, can exactly/approximately be computed \cite{pspin,pedestrians,randommat, 
fyodorov1,realnum1,realnum2,Evnin:2020ddw,Gurau:2020ehg,Kent-Dobias:2020egr,Kent-Dobias2}. 
This makes random tensor models 
 \cite{Ambjorn:1990ge,Sasakura:1990fs,Godfrey:1990dt,Gurau:2009tw,Gurau:2024nzv}
 an attractive arena to study properties of eigenvalues/vectors.
Especially in the limits of large degrees of freedom,
it can be expected that statistical properties converge \cite{pspin,pedestrians,secondmoment}, 
leading to definite statements not depending on each ensemble, as in the thermodynamic limit of statistical physics.
In particular tensor eigenvalue/vector distributions can have sharp edges in such limits, which determine 
the most/best values in applications: e.g., the ground state energy of the spin glass model, 
the largest eigenvalues of random tensors, the best rank-one
approximations of random tensors, the geometric measure of quantum entanglement in quantum information
theory (or injective norms of tensors), etc.
 
Due to the non-linearity, the tensor eigenvalue/vector equations have more varieties \cite{Qi,lim,cart,qibook}
than those of matrices. The most basic are the cases that tensors and eigenvalues/vectors are both real, and such
real cases have been studied in \cite{randommat,secondmoment,Sasakura:2022zwc,
Sasakura:2022iqd,Sasakura:2022axo,Sasakura:2023crd,Kloos:2024hvy,Delporte:2024izt}.  
As for the complex cases, one can consider complex eigenvalues/vectors with tensors being either real or complex. 
There is also the choice of holomorphic equations \cite{Kent-Dobias:2020egr,Kent-Dobias2} 
or non-holomorphic ones \cite{Sasakura:2024awt}, which contain only eigenvectors or both eigenvectors and 
their complex conjugates, respectively. 
Complex cases are important in particular in the applications to quantum information theory, since multipartite
quantum states can be
expressed by complex tensors in general\footnote{For instance $|\Psi\rangle=C_{abc} |a\rangle_1|b\rangle_2|c\rangle_3
\in {\cal H}_1\otimes  {\cal H}_2 \otimes {\cal H}_3$.}. 
In fact the geometric measure of entanglement \cite{shi,barnum,geommeasure,estimate}
(essentially equivalent to the injective norm of the complex random tensor)
in the large degree limit has recently been computed in \cite{Sasakura:2024awt,Dartois:2024zuc} 
by computing the location of the edge of an eigenvalue distribution of 
the complex random tensor \cite{Sasakura:2024awt}.

Because of the difference from the matrix eigenvalue/vector problem, the computation of tensor eigenvalue/vector
distributions require new methods. One of the current authors and the collaborators have recently
computed some of the eigenvalue/vector distributions 
by rewriting them as partition functions of zero-dimensional quantum field theories 
\cite{Sasakura:2022zwc,Sasakura:2022iqd,Sasakura:2022axo,Sasakura:2023crd,
Kloos:2024hvy,Delporte:2024izt,Sasakura:2024awt}.  
This method is systematic and 
powerful, because it can in principle be applied to a wide range of statistical properties of random tensors, and sophisticated quantum field theoretical techniques can be used for exact/approximate computations.  

In this paper, we will apply the quantum field theoretical method to compute three cases of complex eigenvalue/vector 
distributions of real/complex symmetric order-three random tensors. These three cases may be characterized by
the invariance with respect to the Lie-group transformations in the index spaces;  
$O(N,\mathbb{R})$, $O(N,\mathbb{C})$, and $U(N,\mathbb{C})$, respectively\footnote{$N$ denotes
 the dimension of the index space.}.
The eigenvalue/vector equations in the first two cases are holomorphic equations of eigenvectors with random tensors
being real or complex, respectively.
The particular property of these holomorphic cases is that the corresponding quantum field theories are of
fermions with four-fermi interactions (four-fermi theories), containing no bosons. This is an important advantage, 
since such purely fermionic theories are in principle exactly computable \cite{Sasakura:2022zwc,Sasakura:2022iqd,Kloos:2024hvy,Delporte:2024izt,Sasakura:2024awt}. 
The third case is of a non-holomorphic equation of an eigenvector with a random tensor being complex, 
and we would need to introduce bosons as well as fermions to compute the distribution \cite{Sasakura:2022axo}. 
However we will rather compute the ``signed" distribution which counts the distribution with 
a sign factor coming from a Hessian \cite{Sasakura:2022zwc,Kloos:2024hvy,Delporte:2024izt,Sasakura:2024awt}.
The signed distribution can be rewritten as a partition function of a four-fermi theory, which is exactly computable.
In addition it generally agrees with the distribution in the neighborhood of the 
edge \cite{Kloos:2024hvy,Delporte:2024izt,Sasakura:2024awt}, and therefore 
enables us to compute the location of the edge, which is the most important for applications, as mentioned above. 
As an application, we compute the injective norm of the complex symmetric order-three random tensor from the location
of the edge, and obtain good agreement with the numerical result reported in the literature \cite{estimate}.

This paper is organized as follows. In Sections~\ref{sec:complex1}, \ref{sec:complex2}, and \ref{sec:complex3}, 
we compute the three cases of the complex eigenvalue/vector distributions, respectively. 
We obtain the exact closed-form expressions
of the distributions. We take the large-$N$ limits of these expressions, and derive the equations/values of the edges 
and the transition lines/points. We compute the injective norm of the complex symmetric order-three random tensor
from the location of the edge of the last case.
In Section~\ref{sec:monte} we compare our exact expressions with Monte Carlo simulations for 
crosschecks. The last section is devoted to a summary and discussions. 

In Sections~\ref{sec:complex1}, \ref{sec:complex2}, and \ref{sec:complex3} some of the notations are common, 
though they mean different quantities. This is because we want to avoid complications of notations caused by additional
 indices or symbols. For instance there are 
$S_1$ in \eq{eq:s1cp1}, \eq{eq:s1cp2}, and \eq{eq:s1cp3}, and $B$ in \eq{eq:defofB} and 
\eq{eq:banddcp2} (\eq{eq:3Bmatrix}). These common usages cause no problems, 
as far as the notations are confined to each section; 
the computations in each section are independent from those in the other sections, 
while the computational procedures are very common.  

\section{$O(N,\mathbb{R})$ symmetric case}
\label{sec:complex1}
\subsection{Setup}
\label{sec:setupcomplex1}
We will compute the distribution of the complex eigenvectors of the real symmetric random tensor of 
order-three and dimension $N$.
The tenor is denoted by $C_{abc}=C_{bac}=C_{bca} \in \mathbb{R},\ a,b,c=1,2,\cdots,N$. 
In this section a complex eigenvector $v \in \mathbb{C}^N$ of a tensor $C$ is a solution to the following holomorphic 
eigenvector equation,
\[ 
C_{abc}v_b v_c=v_a.
\label{eq:eveqcomplex1}
\]
We assume repeated indices are summed over throughout this paper, unless otherwise stated.
The equation \eq{eq:eveqcomplex1} 
is invariant under the transformation, $C'_{abc}=T_{a}^{a'}T_{b}^{b'}T_{c}^{c'} C_{a'b'c'}$, 
$v'_a=T_{a}^{a'} v_{a'}$, with  $T\in O(N,\mathbb{R})$.
We restrict ourselves to the eigenvectors which satisfy the following condition:
\[
v\hbox{ and }v^*\hbox{ are linearly independent}.
\label{eq:restriction}
\]
The reason for the restriction comes from that the matrix $B$ in 
\eq{eq:defofB} will be required to be non-singular to rewrite the distribution in terms of quantum field theory.
The eigenvector equation \eq{eq:eveqcomplex1}
requires $\re{v}\neq 0$ in the first place, and the restriction \eq{eq:restriction} additionally requires $\im{v}\neq 0$,
which is independent from $\re{v}$.
Here $\re{\cdot},\im{\cdot}$ represent the real and imaginary parts, respectively, and the notations will be
used throughout this paper.

By defining $f_a=v_a-C_{abc}v_b v_c$, the distribution of the eigenvector $v$ for a given tensor $C$ is 
given by
\s[
\rho(v,C)=&\sum_{i=1}^{{\rm \# sol}(C)} \prod_{a=1}^N \delta (v_{Ra}-v_{Ra}^i)\delta (v_{Ia}-v_{Ia}^i) \\
&=|\det M(v,C)|\ \prod_{a=1}^N \delta(f_{Ra})\delta(f_{Ia}) \\
&=\det M(v,C)\ \prod_{a=1}^N \delta(f_{Ra})\delta(f_{Ia}), 
\label{eq:rhovccp1}
\s]
where $v^i\,(i=1,2,\#{\rm sol}(C))$ are all the solutions to \eq{eq:eveqcomplex1}, $v_{Ra}=\re{v_a},v_{Ia}=\im{v_a},
f_{Ra}=\re{f_a},f_{Ia}=\im{f_a}$. 
Here $M(v,C)$ is the $2N\times 2N$ Jacobian matrix associated to the change of the arguments of the delta functions
from the first line to the second, 
\[
M(v,C)=\left(
\begin{array}{cc}
\frac{\partial f_R}{\partial v_R} &\frac{\partial f_R}{\partial v_I} \\
\frac{\partial f_I}{\partial v_R} & \frac{\partial f_I}{\partial v_I}
\end{array}
\right),
\]
where the $N\times N$ block matrices are defined by $\left(\frac{\partial f_R}{\partial v_R}\right)_{ab}
=\frac{\partial f_{Rb}}{\partial v_{Ra}}$, and so on. 
In fact it is more convenient to use complex notations, which are summarized in \ref{app:complex}. 
By using \eq{eq:detcomp}, we obtain
\[
\det M(v,C)=\det \left(
\begin{array}{cc}
\frac{\partial f}{\partial v} &0 \\
0 & \frac{\partial f^*}{\partial v^*}
\end{array}
\right)=\left | \det \frac{\partial f}{\partial v} \right |^2,
\label{eq:mvc}
\]
where $\left(\frac{\partial f}{\partial v}\right)_{ab}=\delta_{ab}-2 C_{abc}v_c$.
Therefore $\det M(v,C)\geq 0$, which approves the transformation from the second to the third line of \eq{eq:rhovccp1}.

The mean distribution of the complex eigenvectors under the random $C$ (a Gaussian randomness)
is given by 
\[
\rho(v)=\left \langle 
\det M(v,C)\ \prod_{a=1}^N \delta(f_{Ra})\delta(f_{Ia}) 
\right \rangle_C,
\label{eq:defrho1}
\]
where $\langle {\cal O} \rangle_C=A^{-1} \int_{\mathbb{R}^{\# C}} dC\, {\cal O}\, e^{-\alpha C^2}$ with $A=\int_{\mathbb{R}^{\# C}} dC\, e^{-\alpha C^2}$, $\alpha$ is a positive number, $C^2=C_{abc}C_{abc}$, 
and $\# C=N(N+1)(N+2)/6$, 
the number of the independent components of $C$.

By using \eq{eq:deltadef} and the formula $\det M=\int d\bar \psi d\psi\, e^{\bar \psi M \psi}$ with a fermion 
pair \cite{zinn}, the distribution \eq{eq:defrho1} can be rewritten as
\[
\rho(v)=\frac{1}{A\, \pi^{2N}} \int dC d\lambda d\bar \psi d\psi d\bar \varphi d\varphi\, e^S,
\label{eq:origrho}
\]
where we have introduced two pairs of fermions, $(\bar \psi, \psi)$ and $(\bar \varphi,\varphi)$, and 
\s[
S&=-\alpha C^2 +I (f \lambda^* +f^* \lambda)+\bar \psi  \frac{\partial f}{\partial v}  \psi +\bar \varphi \frac{\partial f^*}{\partial v^*} \varphi, \\
&= -\alpha C^2 +I (v_a-C_{abc}v_bv_c)\lambda^*_a+I (v^*_a-C_{abc}v^*_bv^*_c)\lambda_a \\
&\hspace{3cm}+\bpsi_a (\delta_{ab}-2 C_{abc}v_c)\psi_b +\bvphi_a (\delta_{ab}-2 C_{abc}v_c^*)\vphi_b
\label{eq:bareaction}
\s]
with $I$ denoting the imaginary unit.
Here in the first line we have suppressed contracted indices, 
such as $f\lambda^*=f_a \lambda_a^*$,
$\bar \psi \frac{\partial f}{\partial v} \psi=\bar \psi_a \frac{\partial f_b}{\partial v_a} \psi_b$, and so on.
Such suppressions are often used throughout this paper, if they do not raise confusions. 

\subsection{Integration over $C$ and $\lambda$}
The integration over $C,\lambda$ in \eq{eq:origrho} 
can straightforwardly be carried out, because they appear at most quadratically
in \eq{eq:bareaction}. After integration over $C$ one obtains
\[
\rho(v)=\frac{1}{\pi^{2N}} \int d\lambda d\bpsi d\psi d\bvphi d\vphi\, e^{S_1},
\label{eq:rho1}
\]
where
\[
S_1= \bpsi_a \psi_a+\bvphi_a \vphi_a+\frac{1}{\alpha} \left(\bpsi \psi v+\bvphi \vphi v^*\right)^2+S_\lambda
\label{eq:s1cp1}
\]
with $S_\lambda$ denoting the terms containing $\lambda$.
Here the third term is a symmetrized product explicitly given by
\s[
&\left(\bpsi \psi v+\bvphi \vphi v^*\right)^2  =\frac{1}{6}
\sum_\sigma \left(\bpsi_{\sigma_a} \psi_{\sigma_b} v_{\sigma_c}+
\bvphi_{\sigma_a} \vphi_{\sigma_b} v_{\sigma_c}^*\right) \left(\bpsi_a \psi_b v_c+\bvphi_a \vphi_b v_c^*\right) \\
&\hspace{0cm}=-\frac{1}{6} \left( 
\bpsi_i\cdot \bpsi_j \psi_i\cdot \psi_j g_{ij} +\bpsi_i\cdot \psi_j \bpsi_j\cdot \psi_i g_{ij} 
+2 \bpsi_i \cdot \psi_j \bpsi_{ji}\psi_{ij} 
+\psi_i \cdot \psi_j \bpsi_{ij} \bpsi_{ji}+\bpsi_i\cdot \bpsi_j \psi_{ij}\psi_{ji}
\right),
\label{eq:phiphiv2}
\s]
where the sum over $\sigma$ is over all the permutations of $a,b,c$.
For the second line we have introduced $v_1=v,\ v_2=v^*,\ \psi_1=\psi,\  
\psi_2=\vphi,\ \bpsi_1=\bpsi,\ \bpsi_2=\bvphi$, and have defined  $\bpsi_{ij}=\bpsi_{ia}v_{ja}$, and so on,
to simplify the expression. Repeated indices $i,j$ are also summed over, unless otherwise stated.

By defining $\lambda_1=\lambda^*,\ \lambda_2=\lambda$, $S_\lambda$ is given by
\[
S_\lambda=-\frac{1}{12 \alpha} \lambda _{i a} B_{ia \, jb} \lambda_{jb}  +I \lambda_{ia} D_{ia},
\]
where the matrix $B$ is defined by
\[
B_{ia\,jb}=(v_i\cdot v_j)^2 \delta_{ab} +2 (v_i\cdot v_j) v_{ja} v_{ib}, 
\label{eq:defofB}
\]
with $v_i\cdot v_j=v_{ia}v_{ja}$, the former two and the latter two index pairs representing the row and 
the column indices,  and 
\[ 
D_{ia}=v_{ia} +\frac{1}{6 \alpha}\sum_{\sigma} v_{ib}v_{ic} \bpsi_{j\sigma_a} \psi_{j \sigma_b} v_{j \sigma_c}.
\label{eq:defofD}
\]
Note that $i,j$ in \eq{eq:defofB} and $i$ in \eq{eq:defofD} are not summed over, because these indices 
are used to define the quantities on the lefthand sides. We will use this convention that indices used to define 
quantities are not summed over, as it is obvious from contexts.

The limitation \eq{eq:restriction} guarantees that $B$ is not singular because of $|v|^2 \neq |v\cdot v|$.
Then by integrating over $\lambda$ in \eq{eq:rho1} we obtain the quantum field theoretical expression of the 
distribution,
\[
\rho(v)=\left( \frac{6 \alpha}{\pi}\right)^N \left( (-1)^N \det B\right)^{-\frac{1}{2}} \int d\bpsi d\psi 
\, e^{S_2},
\label{eq:rhov1cp1}
\]
where 
\[
S_2=\bpsi_{ia}\psi_{ia}+\frac{1}{\alpha} \left(\bpsi \psi v+\bvphi \vphi v^*\right)^2-3 \alpha D_{ia}B^{-1}_{ia,jb} D_{jb}. 
\]
Here the Jacobian between $\lambda_{ia}$ and $\lambda_{Ra}, \lambda_{Ia}$ 
in the integration measure \eq{eq:defofcomplexmeasure} has been taken into account in \eq{eq:rhov1cp1}.

\subsection{Computation of the quantum field theory}
The matrix $B$ in \eq{eq:defofB} can be separated into the matrices
in the subspace spanned by $v_1,v_2$ and in the one transverse to it, 
which are denoted by $\parallel$ and $\perp$, respectively. Note that
the former subspace is two-dimensional (and the latter $N-2$) because of \eq{eq:restriction}.
Then the transverse part can be computed as 
\s[
&D_{ia}B^{\perp-1}_{ ia, jb}D_{jb}\\& \hspace{.5cm}
=-\frac{1}{9\alpha^2} 
\left(
\bpsit_k\cdot \bpsit_l \psi_{ki} \psi_{lj}+\bpsit_l\cdot  \psit_k \bpsi_{ki} \psi_{lj}+\bpsit_k \cdot \psit_l \bpsi_{lj} \psi_{ki}
+ \psit_k\cdot \psit_l \bpsi_{ki} \bpsi_{lj}
\right)g_{ki}g_{2\,ij}^{-1} g_{jl},
\label{eq:DBD}
\s]
where $\bpsit_i$ denotes the projection of $\bpsi_i$ to the transverse subspace, and so on. 
Here we have introduced two $2\times 2$ matrices
$g$ and $g_2$ by $g_{ij}=v_{ia} v_{ja}$ and $g_{2\,ij}=(v_{ia}v_{ja})^2$, and $g_2^{-1}$ in \eq{eq:DBD} 
is the inverse of the matrix $g_2$.

To compute the parallel part, it is convenient to define a new matrix $\tilde B^{-1}$ from 
$B^{\parallel -1}$ by projecting it to $v_i$:
\[
B^{\parallel -1}_{ia\,jb}=\tilde B^{-1}_{ii'\,jj'} v_{i'a} v_{j'b}.
\]
One can prove that $\tilde B^{-1}$ is the inverse matrix of $\tilde B$ defined by
\[
\tilde B_{ii'\,jj'}=g_{2\,ij} g_{i'j'}+2 g_{ij}g_{ij'} g_{i'j}.
\label{eq:defoftB}
\]
Then 
\s[
D_{ia}B^{\parallel -1}_{iajb} D_{jb}&=D_{ia} v_{i'a} \tilde B^{-1}_{ii'jj'} v_{j'b} D_{jb} \\
&=\left(g_{ii'} +\frac{1}{\alpha}F_{ii'} \right)\tilde B^{-1}_{ii'jj'} \left(g_{jj'} +\frac{1}{\alpha}F_{jj'} \right),
\s]
where 
\[
F_{ij}=\frac{1}{3} \left( 
\bpsi_{kj}\psi_{ki}g_{ki}+\bpsi_{ki}\psi_{kj}g_{ki}+\bpsi_{ki}\psi_{ki}g_{kj}
\right).
\]

In the expressions above, the transverse parts of the fermions, $\bpsi^\perp_i$, etc., appear only in the form
of the inner products, $\bpsi_i^\perp \cdot \bpsi_j^\perp$, and so on. To take advantage of this fact
for further computations, let us 
define\footnote{Note that some inner products, such as $\bpsi_1^\perp \cdot \bpsi_1^\perp$, vanish because of 
the anti-commutativity of fermions.}
\[
K^\perp=k_1\, \bpsi^\perp_1 \cdot \bpsi^\perp_2 +k_2\, \psi^\perp_1 \cdot \psi^\perp_2+k_{ij}\, \bpsi^\perp_i\cdot
\psi^\perp_j, 
\]
and consider 
\[
Z^{\perp}(k)=\int d\bpsi^\perp d\psi^\perp \, e^{K^\perp}.
\]
Then the inner products of the transverse fermions can be represented by 
the derivatives of $Z^\perp(k)$ with respect to $k_i,k_{ij}$. For instance, 
\[ 
\int d\bpsi^\perp d\psi^\perp \,  \bpsi_1^\perp \cdot \bpsi_2^\perp \, e^{K^\perp}=\frac{\partial}{\partial k_1} Z^{\perp}(k).
\]
Therefore in \eq{eq:rhov1cp1}
\[
\int d\bpsi d\psi\, e^{S_2}=\left. \int d\bpsi^\parallel d\psi^\parallel \, 
e^{\tilde S (\bar \psi^\parallel,\psi^\parallel,\frac{\partial}{\partial k})} Z^\perp(k)
\right|_{k_{11}=k_{22}=1,\atop k_1=k_2=k_{12}=k_{21}=0},
\label{eq:z2}
\]
where $\tilde S (\bar \psi^\parallel,\psi^\parallel,\frac{\partial}{\partial k})$ is defined by
\[
\tilde S\left( \bar \psi^\parallel,\psi^\parallel,\frac{\partial}{\partial k}\right)
=S_2-\bpsi^\perp_i\cdot\psi^\perp_i
\label{eq:s2psi}
\]
with all the inner products of the transverse fermions being replaced 
by  $\bpsi_1^\perp \cdot \bpsi_2^\perp\rightarrow \frac{\partial}{\partial k_1}$, and so on.
In fact, 
\[
Z^\perp=\left(k_{11}k_{22}-k_{12} k_{21}-k_1k_2\right)^{N-2}
\label{eq:zperpexp}
\]
by explicit computations using the fact that each component of the transverse fermions in $K^\perp$ is independent.

Now the remaining task is to perform the integration over the parallel components of the fermions. To this end
it is convenient to rewrite $\bar\psi^\parallel,\psi^\parallel$
in $\tilde S\left( \bar \psi^\parallel,\psi^\parallel,\frac{\partial}{\partial k}\right)$ in terms of 
the projected fermions $\bar \psi_{ij},\psi_{ij}$.
We perform the replacement $\bar \psi_i\cdot \bar \psi_j=\bar \psi_i^\perp \cdot \bar \psi_j^\perp+
\bar \psi_i^\parallel \cdot \bar \psi_j^\parallel$ in \eq{eq:phiphiv2}, where the parallel term can further be rewritten as 
$\bar \psi_i^\parallel \cdot \bar \psi_j^\parallel=\bar \psi_{ii'} g^{-1}_{i'j'} \bar \psi_{jj'}$,
and so on, by using the identity $\delta_{ab}=v_{ia} v_{jb} g^{-1}_{ij}$ in the parallel subspace.
Then $S\left( \bar \psi^\parallel,\psi^\parallel,\frac{\partial}{\partial k}\right)$ is fully represented 
in terms of $\bar \psi_{ij},\psi_{ij}$ and the derivatives with respect to $k_i,k_{ij}$. 

We can also rewrite the integration measure of the parallel components in \eq{eq:z2} 
in terms of the projected fermions.  By explicitly writing it down, we obtain
\[
d\bar \psi^\parallel d\psi^\parallel=\prod_{i,j=1}^2 d\bar \psi_{ij} d\psi_{ij}\,  (|v|^4-v\cdot v\, v^*\cdot v ^*)^2.
\]

Now we want to perform the explicit integration over $\bar \psi_{ij},\psi_{ij}$.  However, performing this manually is too cumbersome due to the numerous terms in $\tilde S\left( \bar \psi^\parallel,\psi^\parallel,\frac{\partial}{\partial k}\right)$. Therefore, we utilized a Mathematica package for fermionic integration.\footnote{We used grassmann.m, which can be downloaded from 
 \url{https://sites.google.com/view/matthew-headrick/mathematica}.}.  
After explicitly doing this by using the package and putting
the result into \eq{eq:rhov1cp1} (See \ref{app:B} for $\det B$), we obtain
\[
\rho(v)=\left. \frac{1}{3} \left(\frac{6 \alpha}{\pi}\right)^N (b^2-aa^*)^{-\frac{N}{2}-1}(b^2+aa^*)^{-\frac{N}{2}+1}
\exp (f) \,{\cal O}^\parallel\, {\cal O}^\perp Z^\perp \right|_{k_{11}=k_{22}=1,\atop k_1=k_2=k_{12}=k_{21}=0},
\label{eq:rhoderexp}
\]
where 
\s[
&f=
-\alpha \frac{6 b^5 - 6 b^4 ( a + \as) + 2 a  \as b^3 - a^2  \as{}^2 (a + \as) + 
 3 a \as (a + \as)b^2 }{(b^2-a\as)^3}, \\
&{\cal O}^\parallel=
\frac{b^4 + 6 a \as b^2  + a^2 \as{}^2 - 2 b (b^2 + a \as) (a + \as)}{(b^2-a \as)^2}\\
&\hspace{1.cm}+\frac{ aa^* (b^2-a\as)^2}{9 \alpha^2 (b^2+a\as)^2} \left( 
-\frac{\partial^2}{\partial k_1 \partial k_2}
-\frac{\partial^2}{\partial k_{12} \partial k_{21}}+\frac{\partial^2}{\partial k_{11} \partial k_{22}}\right)
\\
&\hspace{1.cm}
-\frac{a(b^2+a\as-2 \as b)}{3 \alpha (b^2+a\as)} \frac{\partial}{\partial k_{11}}
-\frac{\as (b^2+a\as-2 a b)}{3 \alpha (b^2+a\as)} \frac{\partial}{\partial k_{22}}, \\
&{\cal O}^\perp=\exp \left( -\frac{1}{6 \alpha} \left( 2 g_{12} \frac{\partial^2}{\partial k_1 \partial k_2} 
+g_{ij} \frac{\partial^2}{\partial k_{ij} \partial k_{ji}}\right) \right),
\label{eq:opara}
\s]
with the parameters defined 
by\footnote{For notational simplicity we use $a,b$ to represent the inner products of $v,v^*$. 
The same characters are used for the indices of tensors and vectors, but this will not cause any confusions 
from the contexts.}
\s[
&a=g_{11}=v \cdot v,\\
&a^*=g_{22}=v^*\cdot v^*,\\
&b=g_{12}=v \cdot v^*. 
\s]

We can express ${\cal O}^\perp Z^\perp$ in \eq{eq:rhoderexp} more explicitly  by using an identity shown in an appendix of
\cite{Sasakura:2024awt}:
\[
\exp\left( \frac{\partial}{\partial y_i} G_{ij} \frac{\partial}{\partial y_j} \right) (y_iH_{ij} y_j)^n
=n! \left.  \left(\det\left( 1- 4 l H G \right)\right)^{-\frac{1}{2}} \exp \left( 
y \left(1- 4 l  H G \right)^{-1} l H y \right)\right|_{l^n},
\label{eq:theidentity}
\]
where in our case $y=(k_1,k_2,k_{11},k_{12},k_{21},k_{22})$, 
the matrices $G,H$ can be read from $O^\perp$ in \eq{eq:opara} and $Z^\perp$ in \eq{eq:zperpexp}, respectively,
$l$ is an auxiliary expansion parameter, and $F(l)|_{l^n}$ denotes taking the coefficient of the $l^n$ term 
in the series expansion of $F(l)$ around $l=0$. We find
\s[
&{\cal O}^\perp Z^\perp=\left. \frac{\Gamma(N-1)}
{\left(1-\frac{b }{3 \alpha} l \right)^2 \sqrt{1-\frac{ a a^* }{9 \alpha^2}l^2}
}\exp \left(l \left( -\frac{k_1 k_2+k_{12}k_{21}}{1-\frac{b }{3 \alpha} l }+\frac{-\frac{ a^* k_{11}^2 + a k_{22}^2}{6 \alpha} l
 + k_{11} k_{22}}{1 - \frac{ a \as }{9 \alpha^2}l^2}\right)\right)\right|_{l^{N-2}} \\
&= \frac{\Gamma(N-1)}{2 \pi I}
\oint_{{\cal C}_0} 
\frac{dl }
{l^{N-1} \left(1-\frac{b }{3 \alpha} l \right)^2 \sqrt{1-\frac{ a \as }{9 \alpha^2}l^2}
}
\exp \left(l \left( -\frac{k_1 k_2+k_{12}k_{21}}{1-\frac{b }{3 \alpha} l }+\frac{-\frac{ \as k_{11}^2 + a k_{22}^2}{6 \alpha} l
 + k_{11} k_{22}}{1 - \frac{ a \as }{9 \alpha^2}l^2}\right)\right),
 \label{eq:intl}
\s] 
where ${\cal C}_0$ is an anti-clockwise contour around $l=0$. 

$\rho(v)$ depends only on $v\cdot v,v\cdot v^*$. 
Therefore it is more convenient to express it in terms of $\vr,\vi,\theta$,
where $\vr=|\re{v}|,\vi=|\im{v}|$, and $\theta$ is the angle between $\re{v}$ and 
$\im{v}$. Then, considering the volume associated to $d\vr d\vi  d\theta$, we obtain
\[
\rho\left(\vr,\vi,\theta\right)= \frac{4 \pi^{N-\frac{1}{2}}\vr^{N-1} \vi^{N-1} \sin^{N-2}\theta}{\Gamma\left[\frac{N}{2}\right] \Gamma\left[\frac{N-1}{2}\right]} \rho(v).
\label{eq:rhofinal}
\]

\subsection{Large-$N$ asymptotic form}
\label{sec:largencp1}
Now let us discuss the large-$N$ asymptotic form. The most important process of taking the large-$N$ limit is 
to apply the saddle point method to the integral over $l$ in \eq{eq:intl}. 
Since the derivatives with respect to $k_i,k_{ij}$ in  ${\cal O}^\parallel$ will generate some inverse powers 
of $1-b l/(3 \alpha)$ and $1-a a^* l^2/(9 \alpha^2)$ from the exponent in \eq{eq:intl}, 
${\cal O}^\parallel {\cal O}^\perp Z^\perp$ is expressed as a sum of the forms,
\[
\oint_{{\cal C}_0} dl \frac{\text{A polynomial function of } l }{l^{N-1} (1-\frac{b}{3 \alpha}l)^{2+n_1} (1- \frac{a \as}{9\alpha^2} l^2)^{\frac{1}{2}+n_2}} \exp\left( \frac{(1-\frac{a+\as}{6 \alpha} l )l}{ 1-\frac{a\as}{9\alpha^2}l^2} \right),
\label{eq:intsum}
\] 
where $n_1=0,1$ and $n_2=0,1,2$, and we have taken $k_{11}=k_{22}=1,\text{others}=0$, as 
is indicated in \eq{eq:rhoderexp}. 

Now let us assume $v\sim 1/\sqrt{N}$, as this is known to be the proper scaling in the former cases
\cite{randommat,Kloos:2024hvy,Delporte:2024izt,Sasakura:2024awt}. 
This corresponds to $a=\tilde a/N,b=\tilde b/N$, where $\tilde a,\tilde b$ are $\sim O(1)$,
 and we also rescale the integration variable as $l\rightarrow N l$. Then 
 \[
  \eq{eq:intsum} \sim N^{-N}  \oint_{{\cal C}_0} dl 
  \frac{\text{A polynomial function of } l }{(1-\frac{\tb}{3 \alpha}l)^{2+n_1} (1- \frac{\ta \tc}{9\alpha^2} l^2)^{\frac{1}{2}+n_2}} \exp\left( N g(l)  \right),
 \label{eq:intexp}
 \]
 where we have ignored some finite powers of $N$ in the overall factor as subdominant contributions, and
 \[
 g(l)=-\log l +\frac{(1-\frac{\ta+\tc}{6 \alpha} l )l}{ 1-\frac{\ta \tc}{9\alpha^2}l^2}.
\label{eq:gform}
 \]

The saddle point equation $g'(l)=0$ has four solutions. They can conveniently be expressed as
\[
l_{\pm\pm}=\frac{4}{(1\pm x)(1\pm x^*)},
\]
where $x=\sqrt{1-4 \ta/(3 \alpha)}$, and the first and the second lower indices of $l_{\pm\pm}$
represent the four choices of the signs in front of $x$ and $x^*$ on the righthand side, respectively. 
The values of $g(l)$ on the saddle points are given by
\s[
g\left ( l_{\pm\pm}\right)=-\log 4+\frac{1}{1\pm x} +\log(1\pm x)+\frac{1}{1\pm x^*} +\log(1\pm x^*).
\s]

Now let us discuss which of these four saddle points are relevant for our computation. 
For general values of $v$, the saddle points $l_{++},l_{--}$ are on the real axis, while 
$l_{+-},l_{-+}$ are complex. 
The complex saddle points cannot dominate over the real ones in our case, 
because $e^{N g(l_{+-})},e^{N g(l_{-+})}$ are complex and conjugate with each other, and 
any linear combinations of them are oscillatory taking both positive and negative values, 
which contradicts the positivity of $\rho(v)$. As for $l_{++},l_{--}$,
by taking the principal branch of the square root for  $x=\sqrt{1-4 \ta/(3 \alpha)}$, 
we may assume $\re{x}>0$, which leads to
\[
0<l_{++}<3\alpha/|\ta|< l_{--}.
\label{eq:ineql}
\]
Since $g'(l)<0$ for $0<l<l_{++}$,\footnote{Because $f(l)\sim -\log(l)$ for $l\sim+0$, $f'(l)<0$ 
for $l\sim +0$.  As $l$ is increased, $f'(l)<0$ continues until $f'(l)=0$ at $l=l_{++}$.}
 the integration contour around the origin can be deformed to the Lefschetz 
thimble\footnote{See \cite{Witten:2010cx} for the saddle point method using Lefschetz thimbles.}
 ${\cal L}_{++}$ going through $l_{++}$. The saddle point $l_{--}$ is not relevant, because it is hidden 
behind ${\cal L}_{++}$ from the origin. 
Therefore $l_{++}$ is the only saddle point which is relevant in our case. 

\begin{figure}
\begin{center}
\includegraphics[width=7cm]{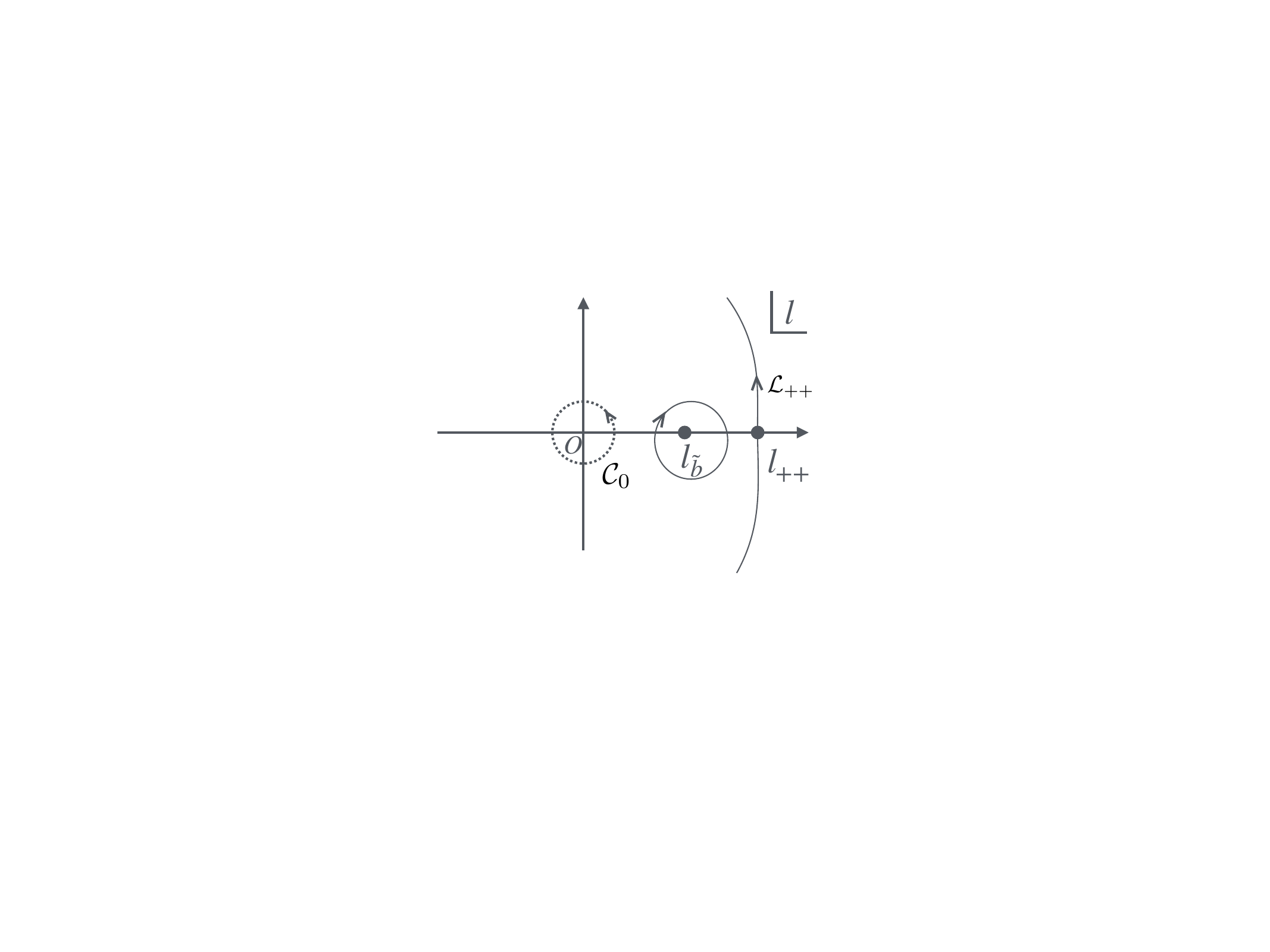}
\caption{
When $l_{\tb} < l_{++}$, the original contour ${\cal C}_0$ (dashed line)
is deformed to the sum of the contour around $l=l_{\tb}$ and 
the Lefschetz thimble ${\cal L}_{++}$ (solid lines) (and other possible subdominant Lefschetz thimbles).}
\label{fig:lb}
\end{center}
\end{figure}

This is not all for this case. We also have to consider the effect of the denominator of \eq{eq:intexp}.  
The factor with $(1-\ta \ta^*/(9 \alpha^2) l^2)$ has no relevance, because $l_{++}<3\alpha/|\tilde a|$ as in \eq{eq:ineql}.
On the other hand, if $1-\tb\, l_{++}/(3 \alpha)<0$,
the deformation of ${\cal C}_0$ to ${\cal L}_{++}$ necessarily generates an extra contribution from the pole at 
$l_{\tb}=3 \alpha/\tb$ (See Figure~\ref{fig:lb}), which is $\sim e^{N g(l_{\tb})}$. Since $0<l_{\tb}<l_{++}$ 
in such a case and $g'(l)<0$ for $0<l<l_{++}$, $g(l_{\tb})>g(l_{++})$.
Therefore we obtain
\[
\eq{eq:intsum}\sim N^{-N} \times \left\{
\begin{array}{ll}
e^{N g(l_{++})},  &\text{for } l_{++}< l_{\tilde b}, \\
e^{N g(l_{\tb})},  &\text{for } l_{++}> l_{\tilde b},
\end{array}
\right.
\]
for large $N$. By also taking into account the other contributions in the leading order of $N$ in \eq{eq:rhoderexp}
and \eq{eq:rhofinal}, which are straightforward to evaluate,   we finally obtain
$\rho\left(\vr,\vi,\theta\right)\sim e^{N h}$ with
\[
h=\tilde f+\log(6\alpha) -\frac{1}{2}\log(\tb^2+\ta \ta^*)+
\left\{
\begin{array}{ll}
g(l_{++}),  &\text{for } l_{++}< l_{\tilde b}, \\
g(l_{\tb}),  &\text{for } l_{++} > l_{\tilde b},
\end{array}
\right.
\label{eq:h}
\]
where $\tilde f$ is obtained by replacing $a\rightarrow \ta,  b\rightarrow \tb$ in $f$. 
Note that the transition at 
$l_{++}=l_{\tb}$ is a continuous transition\footnote{Namely, $h$ and its first derivatives with respect to 
the parameters are continuous.}, because $g(l_{++})=g(l_{\tb})$ and $\frac{\partial}{\partial v} g(l_{\tb})=
\left. \frac{\partial}{\partial v} g(l) \right |_{l=l_{\tb}=l_{++}}=\frac{\partial}{\partial v} g(l_{++})$ due to
$\left. \frac{\partial}{\partial l} g(l) \right |_{l=l_{\tb}=l_{++}}=0$.
This large-$N$ asymptotic expression $h$ has already been obtained by a different method 
in \cite{Kent-Dobias:2020egr,Kent-Dobias2}\footnote{Our expression and theirs can be 
linked by taking $\alpha=1, r=\tilde b/|\tilde a|, \epsilon=1/\sqrt{\tilde a}$.}.

A comment is in order. Since $x=\sqrt{1-4 \ta/(3\alpha)}$, it may be suspected that the singularity and the branch 
cut of the square root may appear in $h$. However, this does not happen. The singularity and the branch cut are 
along $\ta\geq 3\alpha/4$ on the real axis. For $\ta\geq 3\alpha/4$, 
$x$ is pure imaginary and we obtain $l_{++}=3\alpha/\ta$. 
Then $l_{++}\geq l_{\tb}$, because $\tb\geq |\ta|$ in general. This belongs to the second case of \eq{eq:h}, which 
does not depend on $l_{++}$. Therefore $h$ does not contain the square root singularity and the branch cut
coming from $x$.

\begin{figure} 
\begin{center}
\includegraphics[width=7cm]{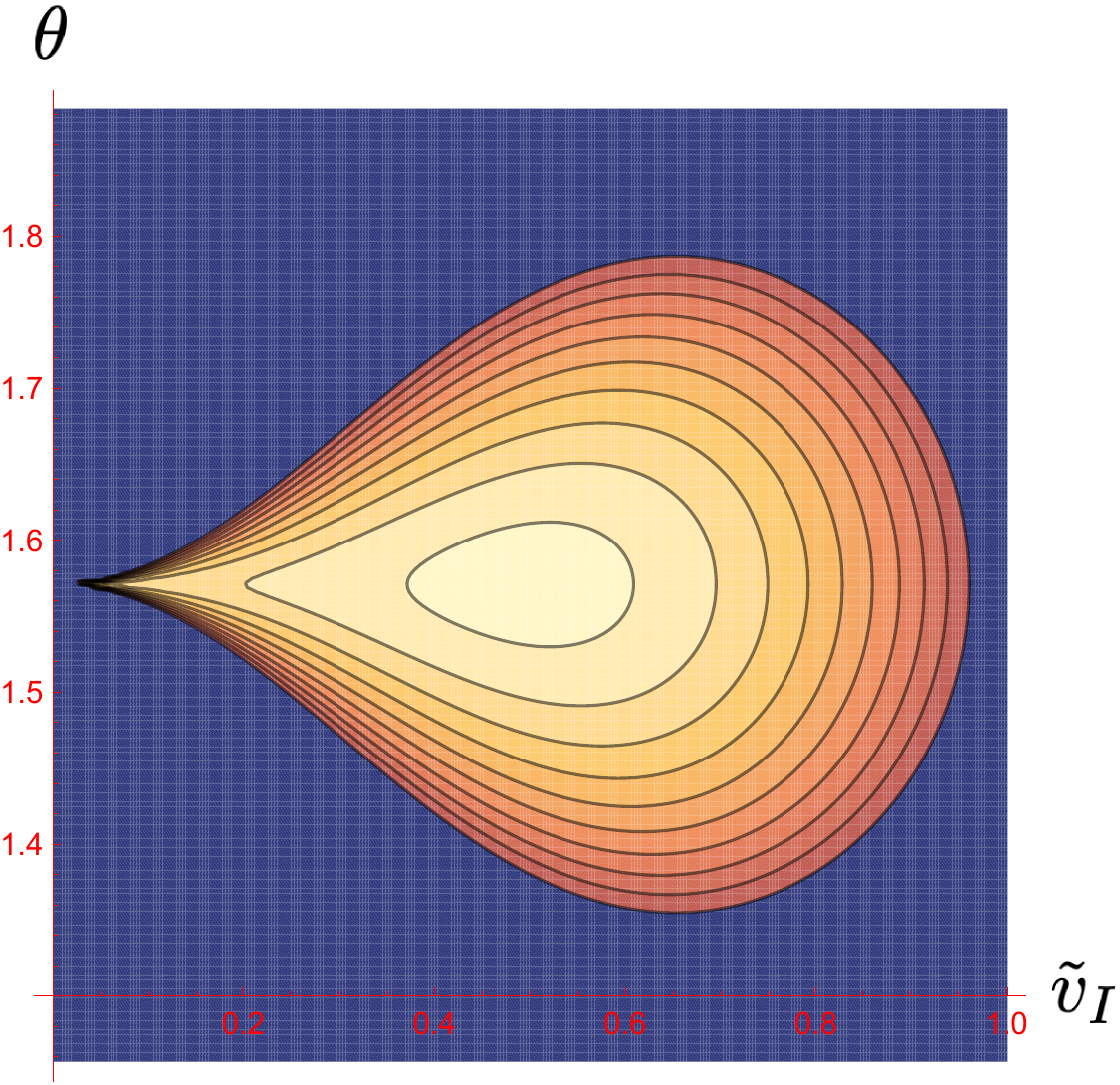}
\caption{A contour plot of $h$ in \eq{eq:h} for $\tvr=1$. The most outer line represents $h=0$, namely the edge
of the large-$N$ limit distribution, while the most inner one $h=0.27$. $h<0$ in the dark region. }
\label{fig:profile}
\end{center}
\end{figure}

\subsection{Large-$N$ profile}
\label{sec:hprofile}
In this subsection we study the profile of $h$ in \eq{eq:h}. We express the parameters in terms of 
$\vr,\vi,\theta$:
\s[
&a=v\cdot v=\vr^2-\vi^2+2 I \vr \vi \cos \theta,\\
&b=v\cdot v^*=\vr^2+\vi^2.
\s] 
We also define the rescaled parameters, $\vr=\tvr/\sqrt{N},\vi=\tvi/\sqrt{N}$.

An example of a contour plot of $h$ is shown in Figure~\ref{fig:profile} for $\tvr=1$.
It seems that the $h>0$ region only touches at $\theta=\pi/2$ in the real value limit $\tvi\rightarrow 0$. 
We will study this aspect in more detail in the following small sections.

\subsubsection{$\tvi\rightarrow0$ limit at $\theta \neq \pi/2$}
Let us study the asymptotic behavior of each term in \eq{eq:h}.
By explicit computation the asymptotic behavior of $\tilde f$ in $\tvi \rightarrow 0$ is 
given by
\[
\tilde f\sim -\frac{\alpha \vr^2 \left(\cos \theta\right)^2}{4\tvi^4 \left(\sin \theta\right)^6 } .
\]

The term $\log(\tb^2+\ta \ta^*)$ in \eq{eq:h} 
is regular for $\tvi\rightarrow 0$. 

There are two possibilities for the last term. Since
the denominator of $g(l_{\tb})$ behaves in $\tvi\rightarrow 0$ as
\[
1-\frac{\ta \ta^*}{9 \alpha^2}l_{\tb}^2 =1-\frac{\ta \ta^*}{\tilde b^2}\sim \frac{4 (\sin\theta)^2 \tvi^2}{\tvr^2},
\]
the asymptotic behavior of $g(l_{\tb})$ is weaker than that of $\tilde f$. As for $g(l_{++})$, there are 
three cases: It is straightforward to show
\s[
1-\frac{\ta \ta^*}{9 \alpha^2} l_{++}^2 \sim
\left \{
\begin{matrix}
\hbox{const.} ,& \tvr<\tilde v_{\rm th},  \\
\sqrt{\tvi}, & \tvr=\tilde v_{\rm th},\\
\tvi, & \tvr>\tilde v_{\rm th},\\
\end{matrix}
\right.
\s]
for $\vi\sim 0$, where $v_{\rm th}=\sqrt{3 \alpha}/2$. 
For all these cases, the asymptotic behavior of $g(l_{++})$ is weaker than that of $\tilde f$. Therefore
for $\vi \sim 0$ at $\theta \neq \frac{\pi}{2}$ 
\[
h \sim -\frac{\alpha \vr^2 \left(\cos \theta\right)^2}{4\tvi^4 \left(\sin \theta\right)^6 } .
\]

\begin{figure} 
\begin{center}
\includegraphics[width=7cm]{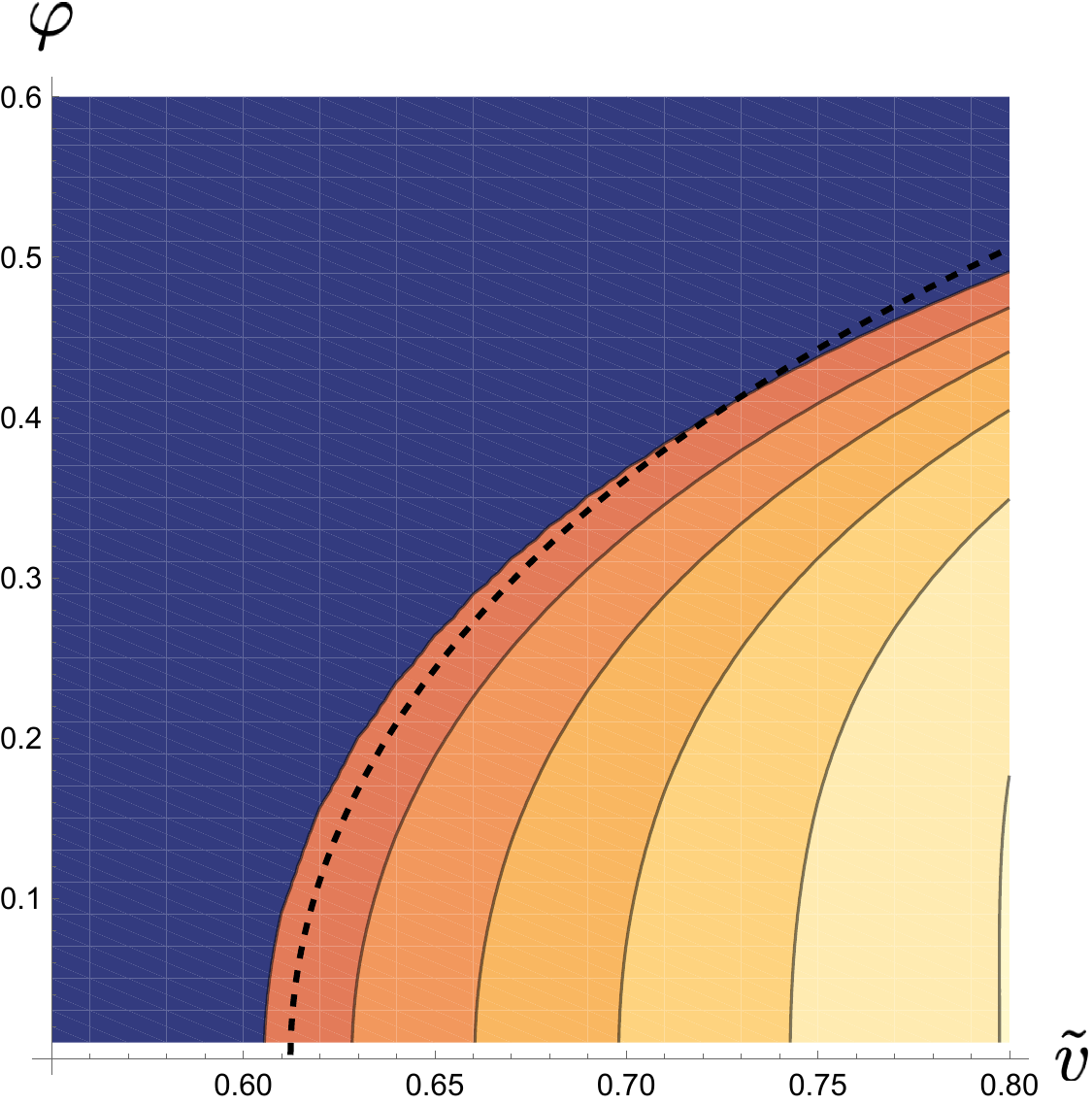}
\caption{A contour plot of $h$ of \eq{eq:hphi}. The most outer line represents $h=0$, while the most inner one $h=0.15$. $h<0$ in the dark region. The dashed line represents the transition line $l_{++}=l_{\tb}$. The edge of 
the distribution ($h=0$) crosses the transition line.}
\label{fig:hphi}
\end{center}
\end{figure}

\subsubsection{At $\theta=\pi/2$}
In this small section we specifically study $h$ for $\theta=\pi/2$ in some detail. Let us parameterize 
$\tvr=\tv \cos \varphi,\tvi=\tv \sin \varphi\ (\tv>0,0< \varphi < \pi/2)$. From a straightforward computation of 
\eq{eq:h}, we obtain
\s[
h_{\theta=\pi/2}=&\log (6 \alpha)-\alpha \frac{7 + 6 \cos(2 \varphi) + \cos(4 \varphi)}{8 \tv^2 (\cos \varphi)^6} 
-\frac{1}{2} \log(\tv^4 (1 +( \cos(2 \varphi))^2))\\
&\hspace{1cm}+
\left\{
\begin{array}{ll}
g(l_{++}),  &\text{for } \cos 2\varphi<f_c \\
g(l_{\tb}),  &\text{for } \cos 2\varphi>f_c
\end{array}
\right.,
\ f_c=-1 + \frac{\sqrt{3 \alpha}}{\tv}.
\label{eq:hphi}
\s]
A contour plot of \eq{eq:hphi} is given in Figure~\ref{fig:hphi}. A peculiar property is that 
the transition line $\cos 2 \varphi=-1 + \sqrt{3 \alpha}/\tv$ crosses the edge of the distribution (namely, $h=0$), and goes outside of the distribution, as has already been noted in \cite{Kent-Dobias:2020egr,Kent-Dobias2}.
This is different from what happens in the real eigenvalue distribution of the real symmetric random tensor \cite{randommat}:
 the transition point is within the distribution and locally stable critical points dominate 
 in the region between the edge and the transition point
  \footnote{The edge of the distribution and the transition point
for the real symmetric random tensor are denoted by $-E_{0}$ and $-E_\infty$, respectively, 
in \cite{randommat}. 
In the region $-E_0 \leq u \leq -E_\infty$, $\Theta_{k=0,p}$ dominates, meaning that 
the critical points with index $k=0$ (locally stable critical points) dominate.  
See \ref{app:aff} for more details.}.

Now let us study the real limit. By taking $\varphi\rightarrow 0$ 
in \eq{eq:hphi}, we obtain
\s[
h_{\theta=\pi/2,\varphi=0}=&\log(6 \alpha)-\frac{7\alpha }{4 \tv^2}-\frac{1}{2}\log(2 \tv^4) \\
&+
\left\{
\begin{array}{ll}
\frac{2}{1+\sqrt{1-4 \tv^2 /(3 \alpha)}}+2 \log(1+\sqrt{1-4 \tv^2 /(3 \alpha)})-2 \log 2,  
&\text{for } \frac{4 \tv^2}{3 \alpha}<1,  \\
\frac{3\alpha}{2 \tv^2}+\log(\frac{4 \tv^2}{3\alpha})-2 \log 2,  &\text{for } \frac{4 \tv^2}{3 \alpha}>1.
\end{array}
\right..
\label{eq:hphizero}
\s]
This expression can be compared with the formulas of the complexity of the critical points
of the spherical $p$-spin spin-glass model derived in \cite{randommat}. 
Some relevant formulas and the mutual relations of the parameters 
are summarized in \ref{app:aff}.  
What is peculiar is that \eq{eq:hphizero} agrees with $\Theta_{p=3}$ for $\frac{4 \tv^2}{3 \alpha}>1$,
but disagrees with it for $\frac{4 \tv^2}{3 \alpha}<1$, instead agreeing with $\Theta_{k=1,p=3}$.  
This implies that the $k=0$ critical points (the locally stable critical points) are missing in \eq{eq:hphizero}.
This phenomenon has already been noted in \cite{Kent-Dobias:2020egr,Kent-Dobias2}. 

\subsection{Absence of optima}
\label{sec:absense}
The eigenvalue/vector equation of the real symmetric random tensor can be regarded as the critical point equation 
of a random potential (i.e. the energy of the $p$-spin spherical model).
The edge of the eigenvalue distribution in the large $N$ limit corresponds to the bottom of the random potential 
\cite{randommat,secondmoment}, and the distribution has the region of the dominance of locally stable 
critical points between the edge and the transition point \cite{randommat}.
A similar structure is expected to exist, if an eigenvalue/vector equation is derived 
from a bounded potential, since most of the critical points near the bound of a potential should be locally stable.
Therefore the peculiar property we encountered in the previous section implies that such a bounded potential
does not exist for the eigenvector equation \eq{eq:eveqcomplex1}. 

To see this more explicitly, let us rewrite our eigenvector equation as a critical point equation of a potential.
A possibility  is
\[
V=\re{C_{abc} w_a w_b w_c}
\]
 with a constraint $w_a w_a=1$. By introducing a Lagrange multiplier $z$, 
 the critical point equation is given by
\[
C_{abc} w_b w_c=z\, w_a 
\]
By comparing with \eq{eq:eveqcomplex1}, one finds that $z=\pm1/\sqrt{v\cdot v}=\pm 1/\sqrt{a}$, and 
$V$ takes
\[
V=\pm {\rm Re} \frac{1}{\sqrt{a}}.
\]
Since $a=v^2(\cos(2 \varphi)+I \sin(2 \varphi) \cos \theta$), $V$ 
diverges at $\varphi=\pi/4,\theta=\pi/2$. 
In fact for large $\tilde v$ with $\varphi=\pi/4,\theta=\pi/2$, the latter case applies in \eq{eq:hphi}, and we obtain 
\[
h\sim \log(2) >0.
\]
Therefore the optimum value is divergent, and the 
corresponding $\tilde v$ is located in the middle of the distribution, not at the edge.  

We would be able to consider another kind of a potential, e.g. $\im{C_{abc} w_a w_b w_c}$ or $| C_{abc} w_a w_b w_c|$. 
But the optimum is again given at $a=0$ with a divergent value.

\section{$O(N,\mathbb{C})$ symmetric case}
\label{sec:complex2}

\subsection{Setup}
We shall look at another alternative case that $C$ is an order-three and complex dimension $N$ symmetric 
random tensor, that is $C_{abc}=C_{bca}=C_{bac} \in \mathbb{C},\ (a,b,c=1,2,\cdots,N)$.
We shall explicitly compute the distribution function ${\rho}(v)$ of the complex eigenvectors satisfying
\begin{equation}
    C_{abc}v_bv_c=v_a,
    \label{eq:eveqcomplex2}
\end{equation}
where $v\in \mathbb{C}^N$.
The equation \eq{eq:eveqcomplex2} 
is invariant under the transformation, $C'_{abc}=T_{a}^{a'}T_{b}^{b'}T_{c}^{c'} C_{a'b'c'}$, 
$v'_a=T_{a}^{a'} v_{a'}$, with  $T\in O(N,\mathbb{C})$.

The eigenvector distribution can be expressed by the same expression as \eq{eq:defrho1} 
with the replacement $f_a=v_a-C_{abc} v_b v_c$ and $\langle {\cal O} \rangle_C=A^{-1} 
\int_{\mathbb{C}^{\# C}} dC \,{\cal O}\, e^{-\alpha C_{abc}^* C_{abc}}$ with 
$A=\int_{\mathbb{C}^{\# C}} dC e^{-\alpha C_{abc}^* C_{abc}}$.
Note that $\det M(v,C)=|\partial f/\partial v|^2\geq 0$ also holds in this case. 
Therefore we obtain a similar expression as \eq{eq:origrho} with some slight changes, 
\s[
S&= -\alpha C^*_{abc}C_{abc} +I (v_a-C_{abc}v_bv_c)\lambda^*_a+I (v^*_a-C^*_{abc}v^*_bv^*_c)\lambda_a \\
&\hspace{3cm}+\bpsi_a (\delta_{ab}-2 C_{abc}v_c)\psi_b +\bvphi_a (\delta_{ab}-2 C^*_{abc}v_c^*)\vphi_b.
\label{eq:bareactioncp2}
\s]

\subsection{Integration over $C$ and $\lambda$}
The Gaussian integration over $C$ can straightforwardly be computed and we obtain the same expression as
\eq{eq:rho1} with a slightly different $S_1$ than \eq{eq:s1cp1},
\[
S_1= \bpsi_a \psi_a+\bvphi_a \vphi_a+\frac{4}{\alpha} \left(\bpsi \psi v\right)\cdot \left(\bvphi \vphi v^*\right)+S_\lambda,
\label{eq:s1cp2}
\]
where the third term is a symmetrized product similar to \eq{eq:phiphiv2}.
Here $S_\lambda$ is given by 
\[
S_\lambda=-\frac{v^4}{3\alpha}\lambda^* B\lambda+i\lambda^*_{a}(v_a+D_a)+i\lambda_a(v_a^*+D_a^*),
\]
where $B$ is a matrix, and $D$, $D^*$ are vectors, given by
\s[
&B_{ab}=\delta_{ab}+\frac{2v^*_a v_b}{\av^2},\\
&D_{a}=\frac{2}{3\alpha}\left(\Bar{\varphi}_{a}\varphi\cdot v \av^2+\Bar{\varphi}\cdot v\varphi_{a}\av^2+\Bar{\varphi}\cdot v\varphi\cdot 
 v v^*_{a}\right),\\
&D_{a}^*=\frac{2}{3\alpha}\left(\Bar{\psi}_{a}\psi\cdot v^* \av^2+\Bar{\psi}\cdot v^* \psi_{a} \av^2+\Bar{\psi}\cdot v^*\psi\cdot v^*v_a\right).
\label{eq:banddcp2}
\s]
We then perform a Gaussian integration over $\lambda$ to give
\[
S_2=\Bar{\psi}\psi+\Bar{\varphi}\varphi+\frac{4}{\alpha}\left(\Bar{\psi}\psi v\right)\cdot \left(\Bar{\varphi}\varphi v^*\right)+W
+\log(A_{\lambda})
\]
where we have
\s[
&A_{\lambda}=\int d\lambda \, e^{-\frac{v^4}{3\alpha}\lambda^*B\lambda},\\
&W=-\frac{3\alpha}{v^4}\left(v^*+D^*\right)B^{-1}\left(v+D\right)
\s]
with $B_{ab}^{-1}=\delta_{ab} -2 v^*_a v_b/ (3 \av^2)$.
We can expand $W$ to obtain
\s[
W=&-\frac{3\alpha}{\av^4}\Big(
\av^2-\frac{2}{3}\frac{v^*\cdot v^*\hskip 0.02in v\cdot v}{\av^2}+v^*\cdot D-\frac{2}{3}\frac{v^*\cdot v^*}{\av^2}\hskip 0.02in v\cdot D+D^*\cdot v-\frac{2}{3}\frac{v\cdot v}{\av^2}\hskip 0.02in v^*\cdot D^*\\
&\hspace{3cm}+D^*\cdot D-\frac{2}{3\av^2} D^*\cdot v^*\, D\cdot v\Big).
\s]
By explicitly computing the inner products, we obtain
\[
W=-\frac{3\alpha}{\av^4}\left(\av^2-\frac{2v^*\cdot v^*\hskip 0.02in v\cdot v}{3\av^2}+W_{2}+W_{4}\right),
\]
where
\s[
&W_{2}=\frac{2}{3\alpha}\Big( \av^2 \Bar{\varphi}\cdot v^*\varphi\cdot v+ \av^2\Bar{\varphi}\cdot v\varphi\cdot v^*+\av^2\Bar{\psi}\cdot v\psi\cdot v^*+ \av^2\Bar{\psi}\cdot v^*\psi\cdot v \\
&\hspace{3cm}-v^*\cdot v^*\Bar{\varphi}\cdot v\varphi\cdot v-v\cdot v\Bar{\psi}\cdot v^*\psi\cdot v^*  \Big), \\
&W_4=\left(\frac{2}{3\alpha}\right)^2\Big(-\av^4\psi\cdot v^*\varphi\cdot v\Bar{\psi}\cdot\Bar{\varphi}+\av^4\psi\cdot v^*\Bar{\varphi}\cdot v\Bar{\psi}\cdot \varphi+\av^4\Bar{\psi}\cdot v^*\varphi\cdot v\psi\cdot\Bar{\varphi}\\
&\hspace{3cm} -\av^4\Bar{\psi}\cdot v^*\Bar{\varphi}\cdot v \psi\cdot \varphi-\av^2\Bar{\psi}\cdot v^*\psi\cdot v^*\Bar{\varphi}\cdot v\varphi\cdot v\Big).
\s]
We now perform the following decomposition:
\s[
    \psi&=\psi_{\perp}+\frac{v}{|v|}\psi_{\parallel}, \ \bar \psi=\bar\psi_{\perp}+\frac{v}{|v|}\bar \psi_{\parallel}, \\
    \varphi&=\varphi_{\perp}+\frac{v^*}{|v|}\varphi_{\parallel}, 
    \ \bar \varphi=\bar\varphi_{\perp}+\frac{v^*}{|v|}\bar \varphi_{\parallel},
    \label{eq:decomppsi}
\s]
where $\psi_{\parallel}=\frac{v^*\cdot \psi}{|v|},\ \psi_{\perp}=\psi-\frac{v}{|v|}\psi_{\parallel}$, 
hence satisfying $v^*\cdot \psi_{\perp}=0$, and so on. 
Then $W_4$ can be rewritten as,
\[
 W_4=\frac{4\av^6}{3\alpha^2}\Bar{\psi}_{\parallel}\psi_{\parallel}\Bar{\varphi}_{\parallel}\varphi_{\parallel}+\frac{4\av^6}{9\alpha^2}\left(-\Bar{\psi}_{\perp}\cdot\Bar{\varphi}_{\perp}\psi_{\parallel}\varphi_{\parallel}+\Bar{\psi}_{\perp}\cdot\varphi_{\perp}\psi_{\parallel}\Bar{\varphi}_{\parallel}+\psi_{\perp}\cdot \Bar{\varphi}_{\perp}\Bar{\psi}_{\parallel}\varphi_{\parallel}-\psi_{\perp}\cdot\varphi_{\perp}\Bar{\psi}_{\parallel}\Bar{\varphi}_{\parallel}\right).
\]
Note that we have the other four-fermi interaction terms from $S_2$: 
$\frac{4}{\alpha}\left(\Bar{\psi}\psi v\right)\cdot \left(\Bar{\varphi}\varphi v^*\right)$. 
By rewriting it with the decomposition \eq{eq:decomppsi}, we obtain
\s[
\frac{4}{\alpha}\left(\Bar{\psi}\psi v\right)\cdot \left(\Bar{\varphi}\varphi v^*\right)&=\frac{4}{6\alpha}\left(\Bar{\psi}\psi v\right)
\cdot \left( \Bar{\varphi}\varphi v^*-\varphi v^*\Bar{\varphi}+v^*\Bar{\varphi}\varphi- \varphi\Bar{\varphi}v^*+\Bar{\varphi}v^*\varphi-v^*\varphi\Bar{\varphi}\right),\\
&=\frac{2\av^2}{3\alpha}\left(-\Bar{\psi}_{\perp}\cdot \Bar{\varphi}_{\perp}\psi_{\perp}\cdot \varphi_{\perp}+\Bar{\psi}_{\perp}\cdot\varphi_{\perp}\psi_{\perp}\cdot \Bar{\varphi}_{\perp}\right)+\frac{4\av^2}{\alpha}\Bar{\psi}_{\parallel}\psi_{\parallel}\Bar{\varphi}_{\parallel}\varphi_{\parallel}\\
&\hspace{1cm}+\frac{4\av^2}{3\alpha}\left(-\Bar{\psi}_{\perp}\cdot \Bar{\varphi}_{\perp}\psi_{\parallel}\varphi_{\parallel}-\psi_{\perp}\cdot \varphi_{\perp}\Bar{\psi}_{\parallel}\Bar{\varphi}_{\parallel}+\Bar{\psi}_{\perp}\cdot \varphi_{\perp}\psi_{\parallel}\Bar{\varphi}_{\parallel}+\psi_{\perp}\cdot \Bar{\varphi}_{\perp}\Bar{\psi}_{\parallel}\varphi_{\parallel}\right).
\label{eq:fourfermisq}
\s]
We notice that $W_4$ cancels the second and the third terms of \eq{eq:fourfermisq} in $S_2$, and therefore
we obtain 
\s[
S_2=&\log (A_{\lambda})-\frac{3\alpha}{\av^2}+\frac{2\alpha \, v\cdot v v^*\cdot v^*}{\av^6}+
\Bar{\psi}_a \left(\delta_{ab}-\frac{2v^*_a v_b}{\av^2}-\frac{2v_a v_b^*}{\av^2}+\frac{2 v\cdot v v_a^* v_b^*}
{\av^4}\right)\psi_b \\
&+\Bar{\varphi}_{a}\left(\delta_{ab}-\frac{2v_a^* v_b}{\av^2}-\frac{2v_a v_b^*}{\av^2}+\frac{2v^*\cdot v^* v_a v_b }
{\av^4}\right)\varphi_{b}+\frac{2\av^2}{3\alpha}\left(-\Bar{\psi}_{\perp}\cdot\Bar{\varphi}_{\perp}\psi_{\perp}\cdot\varphi_{\perp}+\Bar{\psi}_{\perp}\cdot\varphi_{\perp}\psi_{\perp}\cdot\Bar{\varphi}_{\perp}\right).
\label{eq:s2vv}
\s]

\subsection{Computation of the quantum field theory}
To compute the quantum field theory with the action \eq{eq:s2vv}, 
it is necessary to separate the subspace of $v,v^*$ from the rest.
Namely, we consider the decomposition, 
\s[
&\psi=v\psi_1+v^*\psi_2+\psi_{\dperp},\\
&\Bar{\psi}=v\Bar{\psi}_{1}+v^*\Bar{\psi}_{2}+\Bar{\psi}_{\dperp},\\
&\varphi=v^*\varphi_1+v\varphi_2+\varphi_{\dperp},\\
&\Bar{\varphi}=v^*\Bar{\varphi}_{1}+v\Bar{\varphi}_{2}+\Bar{\varphi}_{\dperp},
\s]
where $v\cdot \psi_{\dperp}=v^*\cdot \psi_{\dperp}=0$, and so on.
By explicit computation the Jacobian generated by the transformation of the original integration variables 
to these new variables is given by
\begin{equation}
     \left(v\cdot v v^*\cdot v^*-\av^4\right)^{-2}.
    \label{eq:jacobian}
\end{equation}

Now the remaining task is to explicitly compute the fermionic integration with $S_2$ in \eq{eq:s2vv}. 
Let us first integrate over the fermions in the $v,v^*$ subspace. 
The bilinear term of $\Bar{\psi}, \psi$ in \eq{eq:s2vv} is given by
\[
\Bar{\psi}_a\left(\delta_{ab}-\frac{2v_a^* v_b}{\av^2}-\frac{2v_a v_b^*}{\av^2}+\frac{2 v\cdot v v_a^* v_b^* }{\av^4}\right)\psi_b=\begin{pmatrix}
    \Bar{\psi}_{1} & \Bar{\psi}_{2}
\end{pmatrix}K_1\begin{pmatrix}
    \psi_1\\
    \psi_2
\end{pmatrix}+\Bar{\psi}_{ \perp \mkern-10mu \perp}\cdot\psi_{ \perp \mkern-10mu \perp},
\]
where $K_1$ is
$$K_1=\begin{pmatrix}
    -v\cdot v & -\av^2\\
    -\av^2 & -3v^*\cdot v^*+\frac{2(v^*\cdot v^*)^2v\cdot v}{\av^4}
\end{pmatrix}.
$$
Similarly, we have
$$\Bar{\varphi}_a\left(\delta_{ab}-\frac{2v_a^* v_b}{\av^2}-\frac{2v_a v_b^*}{\av^2}+\frac{2 v^*\cdot v^* v_a v_b }{\av^4}\right)\varphi_b=\begin{pmatrix}
    \Bar{\varphi}_{1} & \Bar{\varphi}_{2}
\end{pmatrix}K_2\begin{pmatrix}
    \varphi_1\\
    \varphi_2
\end{pmatrix}+\Bar{\varphi}_{ \perp \mkern-10mu \perp}\cdot\varphi_{ \perp \mkern-10mu \perp},
$$
where
$$K_2=K_1^*.$$

The four-fermi interaction terms in \eq{eq:s2vv} also contain some components in $v,v^*$ 
directions,  since the $\perp$ subspaces are only transverse to one of $v,v^*$. It is convenient to introduce 
\begin{equation}
    \Tilde{g}=\frac{\av^4-v^*\cdot v^* v\cdot v}{\av^2},\quad g=\frac{2\av^2}{3\alpha},
\end{equation}
and the terms which contain the components of $v,v^*$ in \eq{eq:s2vv} can be summarized as  
\s[
 S_{vv^*}=&
        \Bar{\psi}K_1\psi+\Bar{\varphi}K_2\varphi+ \frac{4\av^2\Tilde{g}^2}{3\alpha}\Bar{\psi}_2\psi_2\Bar{\varphi}_2\varphi_2 \\
        &+\frac{2\av^2\Tilde{g}}{3\alpha}\left(-\Bar{\psi}_2\Bar{\varphi}_2\psi_{ \perp \mkern-10mu \perp}\cdot\varphi_{ \perp \mkern-10mu \perp}-\psi_2\varphi_2\Bar{\psi}_{ \perp \mkern-10mu \perp}\cdot\Bar{\varphi}_{ \perp \mkern-10mu \perp}+\Bar{\psi}_2\varphi_2\psi_{ \perp \mkern-10mu \perp}\cdot\Bar{\varphi}_{ \perp \mkern-10mu \perp}+\psi_2\Bar{\varphi}_2\Bar{\psi}_{ \perp \mkern-10mu \perp}\cdot\varphi_{ \perp \mkern-10mu \perp}\right).
\s]
Then the integration over the $v,v^*$ components can be computed as 
\s[
&\int \prod_{i=1}^2 d\Bar{\psi}_i d\psi_i d\Bar{\varphi}_i d\varphi_i \, e^{S_{vv^*}}\\
&\ \ =
\det  K_1 \det  K_2+K_1^{11}K_2^{11}\left(\frac{4\av^2\Tilde{g}^2}{3\alpha}+\left(\frac{2\av^2\Tilde{g}}{3\alpha}\right)^2\left(-\Bar{\psi}_{ \perp \mkern-10mu \perp}\cdot\Bar{\varphi}_{ \perp \mkern-10mu \perp}\psi_{ \perp \mkern-10mu \perp}\cdot\varphi_{ \perp \mkern-10mu \perp}+\psi_{ \perp \mkern-10mu \perp}\cdot\Bar{\varphi}_{ \perp \mkern-10mu \perp}\Bar{\psi}_{ \perp \mkern-10mu \perp}\cdot\varphi_{ \perp \mkern-10mu \perp}\right)\right),
\label{eq:intpara}
\s]
where, $K^{ij}_{k}$ denote the $ij$-th component of matrix $K_k$. 

The remaining fermionic terms in \eq{eq:s2vv} are of the $\dperp$ components, and are given by
\[
S_{ \perp \mkern-10mu \perp}=\Bar{\psi}_{ \perp \mkern-10mu \perp}\cdot \psi_{ \perp \mkern-10mu \perp}+\Bar{\varphi}_{ \perp \mkern-10mu \perp}\cdot\varphi_{ \perp \mkern-10mu \perp}+g\left(-\Bar{\psi}_{ \perp \mkern-10mu \perp}\cdot\Bar{\varphi}_{ \perp \mkern-10mu \perp}\psi_{ \perp \mkern-10mu \perp}\cdot\varphi_{ \perp \mkern-10mu \perp}+\psi_{ \perp \mkern-10mu \perp}\cdot\Bar{\varphi}_{ \perp \mkern-10mu \perp}\Bar{\psi}_{ \perp \mkern-10mu \perp}\cdot\varphi_{ \perp \mkern-10mu \perp}\right).
\]
To compute the four-fermi theory, we consider
\[
K_{ \perp \mkern-10mu \perp}=\Bar{\psi}_{ \perp \mkern-10mu \perp}\cdot \psi_{ \perp \mkern-10mu \perp}+\Bar{\varphi}_{ \perp \mkern-10mu \perp}\cdot\varphi_{ \perp \mkern-10mu \perp}+k_1\Bar{\psi}_{ \perp \mkern-10mu \perp}\cdot\Bar{\varphi}_{ \perp \mkern-10mu \perp}+k_2\psi_{ \perp \mkern-10mu \perp}\cdot\varphi_{ \perp \mkern-10mu \perp}+k_3\Bar{\psi}_{ \perp \mkern-10mu \perp}\cdot \varphi_{ \perp \mkern-10mu \perp}+k_4\psi_{ \perp \mkern-10mu \perp}\cdot\Bar{\varphi}_{ \perp \mkern-10mu \perp}.
\label{eq:Kperp}
\]
Then
\s[
Z_{\dperp}(g)=&\int d\Bar{\psi}_{\dperp}d\psi_{ \dperp }d\Bar{\varphi}_{ \dperp}d\varphi_{ \dperp}\, e^{S_{ \dperp}} \\
=&\left.e^{g\left(-\frac{\partial}{\partial k_1}\frac{\partial}{\partial k_2}+\frac{\partial}{\partial k_3}\frac{\partial}{\partial k_4}\right)}\int d\Bar{\psi}_{ \perp \mkern-10mu \perp}d\psi_{ \dperp}d\Bar{\varphi}_{ \dperp}d\varphi_{ \dperp}\, e^{K_{ \perp \mkern-10mu \perp}}\right|_{\forall k_i=0} \\
=&\left. e^{g\left(-\frac{\partial}{\partial k_1}\frac{\partial}{\partial k_2}+\frac{\partial}{\partial k_3}\frac{\partial}{\partial k_4}\right)}(1-k_1k_2+k_3k_4)^{N-2}\right|_{\forall k_i=0}.
\s]
By applying the formula \eq{eq:theidentity} to the last expression, we obtain 
\[
    Z_{ \dperp}(g)=(N-2)!\, \left. \left(1-g\, l\right)^{-2}e^l \right|_{l^{N-2}}.
\] 
Now we collect all the multiplicative factors from each of the integrals we have performed and the 
Jacobian factor \eq{eq:jacobian}. The last four-fermi terms in \eq{eq:intpara} can be incorporated 
by taking the derivative $\frac{\partial Z_{\dperp}(g)}{\partial g}$.
Then the final form of $\rho$ is obtained as
\begin{equation}
\begin{aligned}
     \rho(v)=&3^{N-1}\pi^{-N}\alpha^N v^{-4N}\left(N-2\right)!\hskip 0.02in \exp\left(-\frac{3\alpha}{\av^2}+\frac{2\alpha v\cdot vv^*\cdot v^*}{\av^6}\right)\\
    & \times \left.\left(\left(1-\frac{2v\cdot v v^*\cdot v^*}{\av^4}\right
    )^2(1-g\,l)+2g\frac{v\cdot vv^*\cdot v^*}{\av^4}\right)\left(1-g\,l\right)^{-3}e^l\right|_{l^{N-2}}.
    \label{eq:finalrhocp2}
\end{aligned}   
\end{equation}

\subsection{Large-$N$ asymptotic form}
\label{sec:largencp2}
We now perform the large-$N$ analysis of $\rho$. As in Section~\ref{sec:largencp1}, the major task
is to rewrite \eq{eq:finalrhocp2} into an integral form and analyze its saddle points (or Lefschetz thimbles).
In the current case, we will do this for 
\s[
\left. (1-g\,l)^{-n}e^l \right|_{l^{N-2}}&=\frac{1}{2 \pi I} \oint_{{\cal C}_0} dl \, l^{-N+1} (1-g\,l)^{-n}e^l \\
&\sim \frac{N^{-N}}{I} \oint_{{\cal C}_0} dl\, (1-\tilde g\,l)^{-n} e^{N(-\log l +l)},
\label{eq:intformlargencp2}
\s]
where $n=2,3$, and the integral is over a counterclockwise contour around the origin. 
In the second line we have performed the rescaling $g=\tilde g/N$ and 
have changed $l\rightarrow N l$, as in the previous case.

The saddle point of the exponent $f(l)=-\log l +l$ is easily obtained as $l=1$,
and  moreover $\frac{\partial^2 f(l)}{\partial l^2}>0$ at the point, which assures that the original contour 
can be deformed to the Lefschetz thimble going through the saddle point. However, 
as in Section~\ref{sec:largencp1}, for $\tilde g>1$ the deformation generates a contribution from the pole at $l=1/\tilde g$ in \eq{eq:intformlargencp2}. Thus we obtain
\[
\left. (1-g\,l)^{-n}e^l \right|_{l^{N-2}} \sim N^{-N} \times
\left\{
\begin{array}{ll}
e^N & \tilde g<1, \\
e^{N\left(\log \tilde g +\frac{1}{\tilde g}\right)} & \tilde g>1.
\end{array}
\right.
\label{eq:largenintcp2}
\] 

By changing the parameters from $v$ to $\vr,\vi,\theta$ as in \eq{eq:rhofinal}, using \eq{eq:largenintcp2}, and 
taking into account all the factors in \eq{eq:rhofinal} and \eq{eq:finalrhocp2} in the large-$N$ limit, we obtain 
\[
\rho(\vr,\vi,\theta)\sim e^{Nh},
\]
where
\[
h=\log(6 \alpha) +\log(\tilde v_R \tilde v_I \sin\theta)-4 \log|\tilde v|
-\frac{3\alpha}{|\tilde v|^2}+\frac{2\alpha\, \tilde v\cdot \tilde v\, \tilde v^*\cdot \tilde v^*}{|\tilde v|^6}+\left\{
\begin{array}{ll}
1 & \tilde g<1, \\
\log \tilde g +\frac{1}{\tilde g} & \tilde g>1,
\end{array}
\right.
\label{eq:hcp2}
\]
with the notations
$\tilde v=v/\sqrt{N},\ \tilde v_R=|\re{\tilde v}|,\ \tilde v_I=|\im{\tilde v}|,\ \tilde g=\frac{2 |\tilde v|^2}{3 \alpha}$.

\subsection{Large-$N$ profile}
\begin{figure}
\begin{center}
\includegraphics[width=7cm]{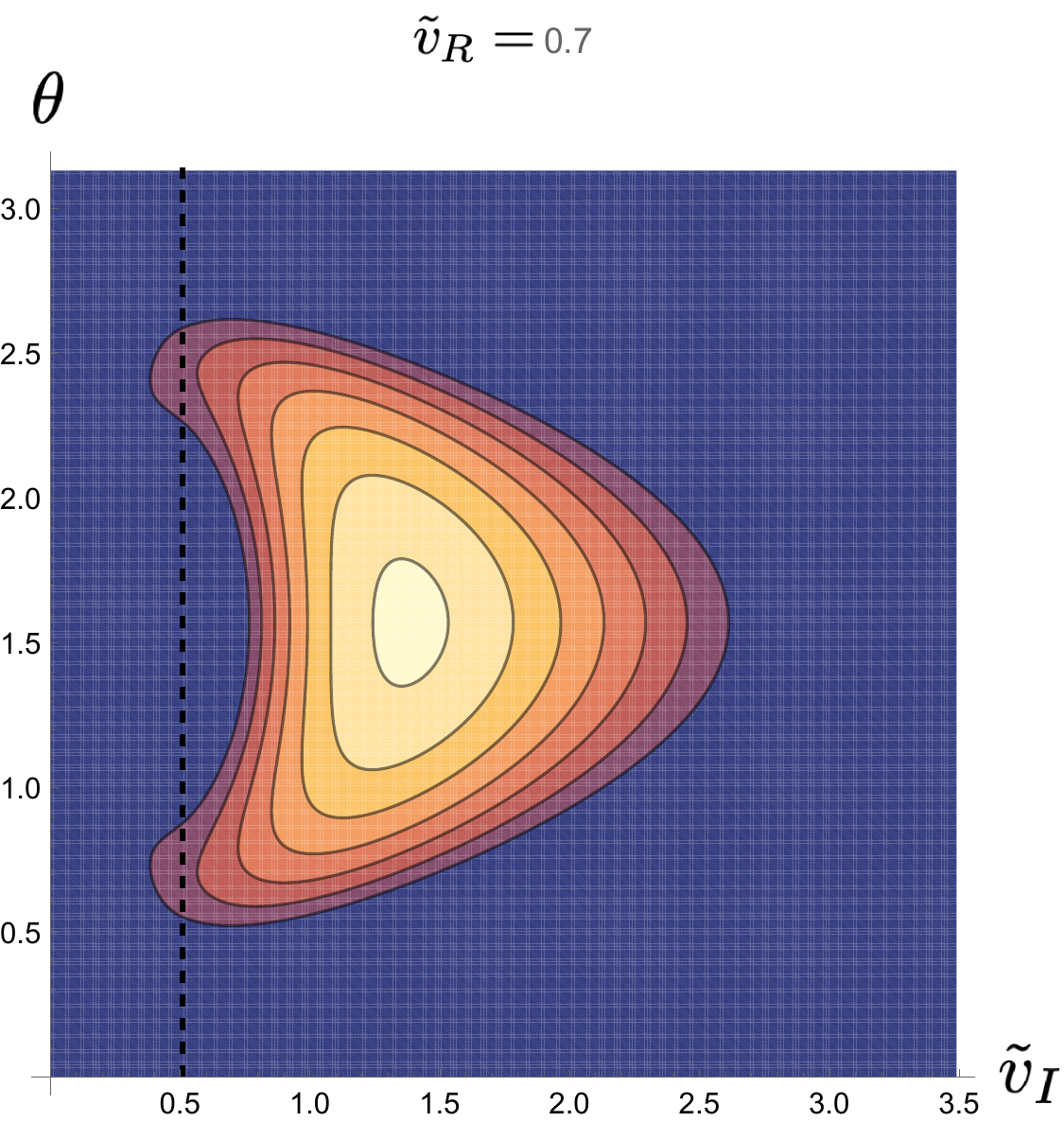}
\hfil
\includegraphics[width=7cm]{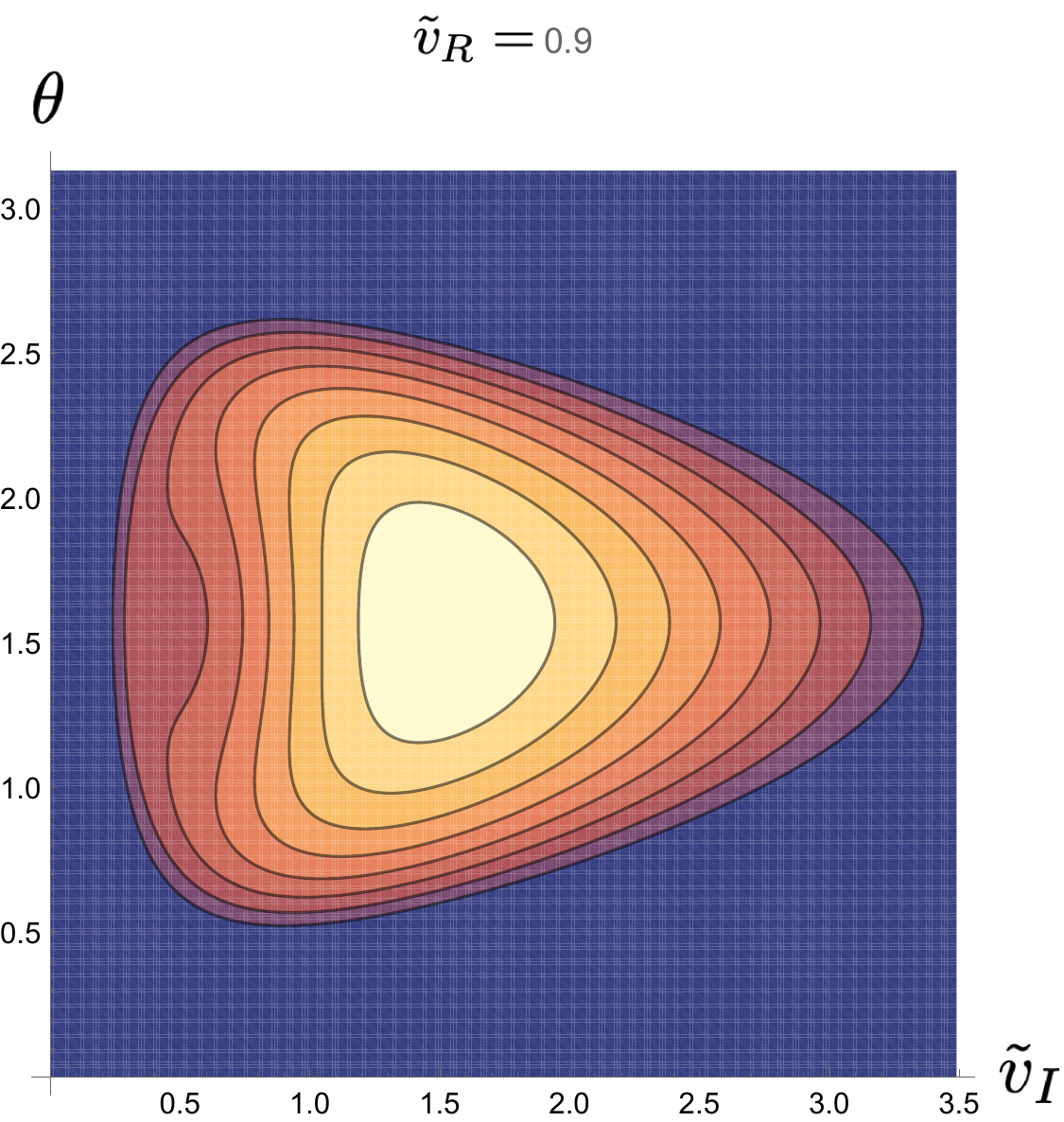}
\caption{Two examples of the contour plots of $h$ in \eq{eq:hcp2} for $\tilde v_R=0.7$ (left) and $\tilde v_R=0.9$ (right),
respectively. The most outer lines represent $h=0$ corresponding to the edges, 
and the most inner lines represent $h=0.3$ (left) and $h=0.35$ (right), respectively.
The dashed line in the left panel is the transition line $\tilde g=1$.}
\label{fig:hprofilecp2}
\end{center}
\end{figure}

In Figure~\ref{fig:hprofilecp2} two examples of the contour plots of $h$ are shown. In the left example, 
the transition line crosses the $h=0$ edge, as was also seen in Section~\ref{sec:hprofile}.
In the right example, the transition line does not exist, because $\tilde v_R ^2> 3 \alpha/2$. 
The situation is similar to the previous case, as discussed in Section~\ref{sec:hprofile} and \ref{sec:absense}, 
and we cannot find a bounded potential which derives the eigenvector equation \eq{eq:eveqcomplex2}.

\section{$U(N,\mathbb{C})$ symmetric case}
\label{sec:complex3}

\subsection{Setup}
We shall look at another alternative case, where $C$ is an order-three and complex dimension $N$ tensor, and is symmetric in their indices, that is:  
$C_{abc}=C_{bca}=C_{cab}\in\mathbb{C}$. 
We consider complex eigenvectors $v\in\mathbb{C}^n$ which satisfy 
\[ 
C_{abc}^*v_bv_c=v_a^*.
\label{eq:eveqcomplex3}
\]
The equation \eq{eq:eveqcomplex3} 
is invariant under the transformation, $C'_{abc}=T_{a}^{a'}T_{b}^{b'}T_{c}^{c'} C_{a'b'c'}$, 
$v'_a=T_{a}^{a'} v_{a'}$, with  $T\in U(N,\mathbb{C})$.

By defining $f_a=v_a-C_{abc}v^*_b v^*_c$, the distribution of the eigenvector $v$ for a given tensor $C$ is 
given by
\s[
\rho(v,C)=&\sum_{i=1}^{{\rm \# sol}(C)} \prod_{a=1}^N \delta (v_{Ra}-v_{Ra}^i)\delta (v_{Ia}-v_{Ia}^i) \\
&=|\det M(v,C)|\ \prod_{a=1}^N \delta(f_{Ra})\delta(f_{Ia}),
\label{eq:rhovccp3}
\s]
where the Jacobian matrix is given by (in block matrix form)
\[M(v,C)=\begin{pmatrix}
    \mathcal{I} & -2Cv^*\\
    -2C^*v & \mathcal{I}
\end{pmatrix}\label{eq:3matrix}\]
with $\mathcal{I}$ being the $N\times N$ identity matrix.
Since the determinant of $M(v,C)$ is not positive semi-definite in this case, we shall work with 
the signed distribution \cite{Sasakura:2022zwc,Kloos:2024hvy}, where taking the modulus is ignored:
\s[
\rho_{\rm signed}(v,C)=\det M(v,C)\ \prod_{a=1}^N \delta(f_{Ra})\delta(f_{Ia}).
\label{eq:rhosigncvvcp3}
\s]
Like the previous cases, we average the signed distribution function over the random tensor $C$, 
and rewrite the terms in the integrand:
using the definition of the determinant in terms of fermionic variables and replacing the Dirac delta distribution 
in it's integral form make the signed distribution as
\[\rho_{\rm signed} (v)=(\pi)^{-2N}A^{-1}\int d\Bar{\psi}\hskip 0.02in d\psi \hskip 0.02in { d\varphi\hskip 0.02in d\Bar{\varphi}\hskip 0.02in} d\lambda \hskip 0.02in  dC e^{S}, \]
where  $A=\int_{\mathbb{C}^{\# C}} dC e^{-\alpha C_{abc}^* C_{abc}}$, and
\[
S=-\alpha C^*C+\Bar{\psi}\psi-2C\Bar{\psi}\Bar{\varphi}v^*-2C^*\varphi\psi v+\varphi\Bar{\varphi}+I\lambda^{*}\left(v-Cv^*v^*\right)+I\lambda\left(v^*-C^*vv\right)
\label{eq:Sofcp3}
\]
with $C\Bar{\psi}\Bar{\varphi}v^*=C_{abc}\Bar{\psi}_a\Bar{\varphi}_bv_c^*$, and so on.
Note that we have interchanged the convention of $\varphi,\Bar{\varphi}$
compared with the previous cases for later convenience.

As explicitly studied thoroughly in \cite{Kloos:2024hvy}
for the real eigenvalue/vector distribution of the real symmetric random tensor, 
the signed distribution is still very useful, since it can determine
the locations of the edge and the transition point of the genuine distribution, which are vital for applications.
This arises from the region between the edge and the transition point, where locally stable critical points dominate and the two distributions coincide.
Although it has not been proven whether similar properties hold for the present complex case, we will present some 
supporting evidences at the end of Section~\ref{sec:largeNcp3}.

\subsection{Integration over $C$ and $\lambda$}
We now perform the integration over the tensor $C$.
We collect all terms containing $C$ in \eq{eq:Sofcp3}:
\[
S_C=-\alpha C^*C-C\left(2\Bar{\psi}\Bar{\varphi}v^*+I\lambda^*v^*v^*\right)-C^*\left(2\varphi\psi v+I\lambda vv\right).
\]
After performing the Gaussian integration over $C$, the normalization factor $A$ is cancelled, and the overall action 
is obtained as 
\[
S_1=\Bar{\psi}\psi+\varphi\Bar{\varphi}+S_\lambda+\frac{4}{\alpha}\left(\Bar{
\psi
}\Bar{\varphi}v^*\right)\cdot \left(\varphi\psi v\right),
\label{eq:s1cp3}
\]
where the last term is a symmetrized product as in \eq{eq:phiphiv2}, and 
$S_{\lambda}$ is the part containing $\lambda$, which is given by
\s[
 S_\lambda&=-\frac{1}{\alpha}\left(\lambda^*v^*v^*\right)\cdot\left(\lambda vv\right)+\frac{2I}{\alpha}\left(\lambda vv\right)\cdot\left(\Bar{\psi}\Bar{\varphi}v^*\right)+\frac{2I}{\alpha}\left(\lambda^*v^*v^*\right)\cdot\left(\varphi\psi v\right)+I\lambda^*v+I\lambda v^* \\
&=-\frac{v^4}{3\alpha}\lambda^*B\lambda+I\lambda^*\left(v+D\right)+I\lambda\left(v^*+D^*\right).
\s]
Here $B$ is a matrix and $D,D^*$ are vectors, which are given by
\s[
    &B_{ab}=\delta_{ab}+\frac{2v_{a}v^*_{b}}{|v|^2},\\
&D_{a}=\frac{2}{3\alpha}\left(\varphi_{a}\left(\psi\cdot v^*\right)|v|^2+\left(\varphi\cdot v^*\right)\psi_{a} |v|^2+\varphi\cdot v^*\psi\cdot v^* v_{a}\right), \\
&D_{a}^*=\frac{2}{3\alpha}\left(\bpsi_{a}\left(\bvphi\cdot v\right)|v|^2+\left(\bpsi \cdot v\right)\bvphi_{a} |v|^2+\bpsi\cdot v\bvphi \cdot v v^*_{a}\right).
\label{eq:3Bmatrix}
\s]
We then perform a Gaussian integration over $\lambda$, to give us a multiplicative factor of $\pi^N$ and yield the action, 
\[
S_{2}=-\log\left(\det\left(\frac{|v|^4}{3\alpha}B\right)\right)+\Bar{\psi}\psi+\varphi\Bar{\varphi}+\frac{4}{\alpha}\left(\Bar{\psi}\Bar{\varphi}v^*\right)\cdot \left(\varphi\psi v\right)-\frac{3\alpha}{v^4}\left(v^*+D^*\right)B^{-1}\left(v+D\right),
\label{eq:eff3}\]
where the inverse of $B$ in \eq{eq:3Bmatrix} can be simply obtained as
\begin{equation}
    B^{-1}=\mathcal{I}-\frac{2v\otimes v^*}{3|v|^2}=\frac{1}{3}\frac{v\otimes v^*}{|v|^2}+\left(\mathcal{I}-\frac{v\otimes v^*}{|v|^2}\right)=\frac{1}{3}I_{\parallel}+I_{\perp}.
\end{equation}
Here ${\cal I}_{ab}=\delta_{ab}$ and $I_{\parallel},I_{\perp}$ are the projection operators given by
$$I_{\parallel}=\frac{v\otimes v^*}{|v|^2},\quad I_{\perp}=\mathcal{I}-I_{\parallel}.$$
The determinant of $B$ in \eq{eq:3Bmatrix} can be easily evaluated to be equal to 3. We now perform the following decomposition:
\s[
    \varphi&=\left(\mathcal{I}-\frac{v\otimes v^*}{|v|^2}\right)\varphi+\frac{v\otimes v^*}{|v|^2}\varphi=\varphi_{\perp}+\frac{v}{|v|}\varphi_{\parallel},\\
    \bar  \varphi&=\left(\mathcal{I}-\frac{v^*\otimes v}{|v|^2}\right)\bar \varphi+\frac{v^*\otimes v}{|v|^2}\bar \varphi=
    \bar \varphi_{\perp}+\frac{v^*}{|v|}\bar \varphi_{\parallel},
\label{eq:3phi}
\s]
where $\varphi_{\parallel}=\frac{v^*\cdot \varphi}{|v|}$,  $\varphi_{\perp}=\varphi-\frac{v}{|v|}\varphi_{\parallel}$, and so on,
which satisfy $v^*\cdot\varphi_{\perp}=v\cdot\Bar{\varphi}_{\perp}=0$. Similarly, we introduce
\s[
     \psi&=\left(\mathcal{I}-\frac{v\otimes v^*}{|v|^2}\right)\psi+\frac{v\otimes v^*}{|v|^2}\psi=\psi_{\perp}+\frac{v}{|v|}\psi_{\parallel}, \\
      \bar \psi&=\left(\mathcal{I}-\frac{v^*\otimes v}{|v|^2}\right)\bar \psi+\frac{v^*\otimes v}{|v|^2}\bar \psi
      =\bar \psi_{\perp}+\frac{v^*}{|v|}\bar \psi_{\parallel}.
\label{eq:3psi}\s]
Using the decompositions, we can write the vector $D$ in \eq{eq:3Bmatrix} as
\[
D=D_{\perp}+\frac{v}{|v|}D_{\parallel},
\label{eq:96}
\]
where the terms in \eq{eq:96} can be written as
\[
D_{\perp}=\frac{2|v|^3}{3\alpha}\left(\varphi_{\perp}\psi_{\parallel}+\varphi_{\parallel}\psi_{\perp}\right),\quad
D_{\parallel}=\frac{v^*\cdot D}{|v|}=\frac{2|v|^3}{\alpha}\varphi_{\parallel}\psi_{\parallel}.
\]
Similarly, we also have
\[D_{\perp}^{*}=\frac{2|v|^3}{3\alpha}\left(\Bar{\psi}_{\perp}\Bar{\varphi}_{\parallel}+\Bar{\psi}_{\parallel}\Bar{\varphi}_{\perp}\right),\quad 
D_{\parallel}^{*}=\frac{2|v|^3}{\alpha}\Bar{\psi}_{\parallel}\Bar{\varphi}_{\parallel}.
\]
We then compute the following:
\s[
&\left(v^*+D^*\right)B^{-1}\left(v+D\right)=\left(v^*+\frac{v^*D^*_{\parallel}}{|v|}+D^*_{\perp}\right)\left(\frac{1}{3}I_{\parallel}+I_{\perp}\right)\left(v+\frac{v}{|v|}D_{\parallel}+D_{\perp}\right)\\
&\hspace{.5cm}=\frac{1}{3}\left(v^*+\frac{v^*D^*_{\parallel}}{|v|}\right)\left(v+\frac{vD_{\parallel}}{|v|}\right)+D^*_{\perp}\cdot D_{\perp} \\
&\hspace{.5cm}
=\frac{|v|^2}{3}\left(1+\frac{2|v|^2}{\alpha}\Bar{\psi}_{\parallel}\Bar{\varphi}_{\parallel}\right)\left(1+\frac{2|v|^2}{\alpha}\varphi_{\parallel}\psi_{\parallel}\right)+\left(\frac{2|v|^3}{3\alpha}\right)^2\left(\Bar{\psi}_{\perp}\Bar{\varphi}_{\parallel}+\Bar{\psi}_{\parallel}\Bar{\varphi}_{\perp}\right)\cdot \left(\varphi_{\perp}\psi_{\parallel}+\varphi_{\parallel}\psi_{\perp}\right).
\label{eq:100}
\s]
We now consider another term in $S_2$, using the anti-commuting property of the fermionic variables:
\s[
\frac{4}{\alpha}\left(\Bar{\psi}\Bar{\varphi}v^*\right)\cdot\left(\varphi\psi v\right)
&=\frac{2}{3\alpha}\left(\Bar{\psi}\Bar{\varphi}v^*\right)\cdot \left(
\varphi \psi v-\psi v \varphi +v \varphi \psi-\psi \varphi v+\varphi v \psi -v \psi \varphi
\right) \\
&=\frac{2}{3\alpha}\big(-\Bar{\psi}\cdot\varphi\Bar{\varphi}\cdot\psi v^2+\Bar{\psi}\cdot\psi\Bar{\varphi}\cdot v v^*\cdot\varphi+\Bar{\psi}\cdot v\Bar{\varphi}\cdot\varphi\psi\cdot v^* \\
&\hspace{1.5cm}+\Bar{\psi}\cdot\psi\Bar{\varphi}\cdot\varphi v^2-\Bar{\psi}\cdot\varphi\Bar{\varphi}\cdot vv^*\cdot\psi-\Bar{\psi}\cdot v\Bar{\varphi}\cdot\psi v^*\cdot \varphi\big)\\
&=\frac{2|v|^2}{3\alpha}\big(-\Bar{\psi}_{\perp}\cdot\varphi_{\perp}\Bar{\varphi}_{\perp}\cdot\psi_{\perp}+\Bar{\psi}_{\perp}\cdot\psi_{\perp}\Bar{\varphi}_{\perp}\cdot\varphi_{\perp}
+6\Bar{\psi}_{\parallel}\psi_{\parallel}\Bar{\varphi}_{\parallel}\varphi_{\parallel} \\
&\hspace{1.5cm}-2\Bar{\psi}_{\perp}\cdot\varphi_{\perp}\Bar{\varphi}_{\parallel}\psi_{\parallel} 
-2\Bar{\varphi}_{\perp}\cdot\psi_{\perp}\Bar{\psi}_{\parallel}\varphi_{\parallel}
+2\Bar{\psi}_{\perp}\cdot\psi_{\perp}\Bar{\varphi}_{\parallel}\varphi_{\parallel}
+2\Bar{\varphi}_{\perp}\cdot\varphi_{\perp}\Bar{\psi}_{\parallel}\psi_{\parallel}\big),
\label{eq:102}
\s]
where we have used the decomposition of \eq{eq:3phi} and \eq{eq:3psi} for the last line.

We now put \eq{eq:100}, \eq{eq:102} into \eq{eq:eff3}, which leads to the cancellation of the four-fermi interactions among
parallel components and all the mixed terms involving both parallel and transverse components:
\s[
&S_2=-\log \left(\det\left(\frac{|v|^4B}{3\alpha}\right)\right)-\frac{\alpha}{|v|^2}\\
&\hspace{1cm}+\Bar{\psi}\psi+\varphi\Bar{\varphi}-2\varphi_{\parallel}\psi_{\parallel}-2\Bar{\psi}_{\parallel}\bar \varphi_{\parallel}
+\frac{2|v|^2}{3\alpha}\left(-\Bar{\psi}_{\perp}\cdot\varphi_{\perp}\Bar{\varphi}_{\perp}\cdot\psi_{\perp}+\Bar{\psi}_{\perp}\cdot\psi_{\perp}\Bar{\varphi}_{\perp}\cdot\varphi_{\perp}\right).
\label{eq:103}
\s]

\subsection{Computation of the quantum field theory}
Since the parallel and the transverse components are decoupled in \eq{eq:103}, they can be computed separately. 
By expanding the exponent, we can straightforwardly compute the parallel part as
\[
\int d\bpsi_\parallel d\psi_\parallel d\vphi_\parallel d\bvphi_\parallel \, e^{\Bar{\psi}_\parallel \psi_\parallel+\varphi_\parallel\Bar{\varphi}_\parallel
-2\varphi_{\parallel}\psi_{\parallel}-2\Bar{\psi}_{\parallel}\bar\varphi_{\parallel}}=-3.
\]
Note that, as commented below \eq{eq:Sofcp3}, the order of the integration measure of $\vphi,\bvphi$ 
is opposite compared to $\bpsi,\psi$ for convenience, which has lead to the overall minus sign.

Collecting all the computations so far, we obtain
\[
 \rho_{\rm signed} (v)=-3^{N}\hskip 0.02in\pi^{-N}\hskip 0.02in\alpha^N \hskip 0.02in |v|^{-4N}\hskip 0.02in 
 e^{-\alpha/|v|^2}\hskip 0.02in Z_{\perp},
\label{eq:tildemul}\]
where $Z_{\perp}$ is the partition function of the transverse fermions.
This is given by
\[
    Z_{\perp}=\int d\Bar{\psi}_{\perp}\hskip 0.02in d\psi_{\perp} \hskip 0.02in
    d \varphi_{\perp}\hskip 0.02in d\Bar{\varphi}_{\perp} \hskip 0.02in e^{\mathcal{S}_{\perp}},
\label{eq:perpz3}
\]
where
\[
    S_{\perp}=\Bar{\psi}_{\perp}\cdot\psi_{\perp}-\Bar{\varphi}_{\perp}\cdot\varphi_{\perp}+g\left(-\Bar{\psi}_{\perp}\cdot\varphi_{\perp}\Bar{\varphi}_{\perp}\cdot\psi_{\perp}+\Bar{\psi}_{\perp}\cdot\psi_{\perp}\Bar{\varphi}_{\perp}\cdot\varphi_{\perp}\right)
    \label{eq:sperpcp3}
\]
with
\[
g=\frac{2|v|^2}{3\alpha}.
\label{eq:3g}
\]
Note that the integration measure of $\bvphi_\perp,\vphi_\perp$ in \eq{eq:perpz3}
has an opposite order compared to $\bpsi_\perp,\psi_\perp$, as commented below \eq{eq:Sofcp3}.

The four-fermi theory \eq{eq:perpz3} with \eq{eq:sperpcp3} can be computed by the similar procedure as 
in the previous sections. 
Let us define
\s[
K_\perp&=k_1\Bar{\psi}_{\perp}\cdot\psi_{\perp}+k_2\Bar{\varphi}_{\perp}\cdot\varphi_{\perp}+k_3\Bar{\psi}_{\perp}\cdot\varphi_{\perp}+k_4\Bar{\varphi}_{\perp}\cdot\psi_{\perp} \\
&=\begin{pmatrix}
    \Bar{\psi}_{\perp a} &  \Bar{\varphi}_{\perp a}
\end{pmatrix}\begin{pmatrix}
    k_{1} & k_3\\
    k_4 & k_2
\end{pmatrix}\begin{pmatrix}
      {\psi}_{\perp a } \\
    {\varphi}_{\perp a}
\end{pmatrix}.
\label{eq:Kperpcp3}
\s]
Then \eq{eq:perpz3} can be computed as
\s[
    Z_{\perp}&=\left.\exp\left(g\left(\frac{\partial}{\partial k_1}\frac{\partial}{\partial k_2}-\frac{\partial}{\partial k_3}\frac{\partial}{\partial k_4}\right)\right)\int d\Bar{\psi}_{\perp}\hskip 0.02in d\psi_{\perp} d\varphi_{\perp} d\Bar{\varphi}_{\perp}\hskip 0.02in\hskip 0.02in  \hskip 0.02in e^{K_\perp}\right|_{k_1=1,k_2=-1,k_3=k_4=0} \\
    &=\left.\exp\left(g\left(\frac{\partial}{\partial k_1}\frac{\partial}{\partial k_2}-\frac{\partial}{\partial k_3}\frac{\partial}{\partial k_4}\right)\right)\int (-1)^{N-1}d\Bar{\psi}_{\perp}\hskip 0.02in d\psi_{\perp} d\Bar{\varphi}_{\perp}\hskip 0.02in\hskip 0.02in d\varphi_{\perp} \hskip 0.02in e^{K_\perp}\right|_{k_1=1,k_2=-1,k_3=k_4=0} \\
    &=\left.\exp\left(g\left(\frac{\partial}{\partial k_1}\frac{\partial}{\partial k_2}-\frac{\partial}{\partial k_3}\frac{\partial}{\partial k_4}\right)\right)\left(-k_1k_2+k_3k_4\right)^{N-1}\right|_{k_1=1,k_2=-1,k_3=k_4=0},
    \label{eq:zperpcp3}
\s]
where, from the first line to the second, the order of $\varphi,\bar \varphi$ is interchanged, producing the 
factor $(-1)^{N-1}$. 

Now let us apply \eq{eq:theidentity} to \eq{eq:zperpcp3}, where
\s[
y=\begin{pmatrix}
     k_1 & k_2 & k_3 & k_4
 \end{pmatrix},\ 
 G= g H,\ 
 H=\frac{1}{2}\begin{pmatrix}
     0&-1&0&0\\
     -1&0&0&0\\
     0&0&0&1\\
     0&0&1&0
 \end{pmatrix}.
 \s]
Then we obtain
\s[
Z_{\perp}=(N-1)!\left.\left(1+lg\right)^{-2}\hskip 0.02in \exp\left(\frac{l}{1+lg}\right)\right|_{l^{N-1}}.
\label{eq:zperpfinal}
\s]
Putting \eq{eq:zperpfinal} into \eq{eq:tildemul}, we obtain 
\[
 \rho_{\rm signed} (v)=- 3^{N} \hskip 0.02in\pi^{-N}\hskip 0.02in\alpha^N \hskip 0.02in |v|^{-4N}\hskip 0.02in e^{-\alpha/|v|^2}\hskip 0.02in (N-1)!\left.\left(1+lg\right)^{-2}\hskip 0.02in \exp\left(\frac{l}{1+lg}\right)\right|_{l^{N-1}}.
 \label{eq:rhosemifinalcp3}
\]

Since \eq{eq:rhosemifinalcp3} depends only on $|v|$, it is more useful to consider the distribution as a function 
of $|v|$. The volume factor associated to this change of the variables is that of the $2N-1$ dimensional sphere 
of radius $|v|$, since $|v|^2=|\re v|^2+|\im v|^2$. 
Multiplying this factor $2 \pi^N |v|^{2 N-1}/ \Gamma(N)$ to \eq{eq:rhosemifinalcp3}, we obtain
\[
\rho_{\rm signed}(|v|)=- 2\cdot  3^{N} \alpha^N  |v|^{-2N-1} e^{-\alpha/|v|^2} 
\left.\left(1+l g\right)^{-2}\hskip 0.02in \exp\left(\frac{l}{1+lg}\right)\right|_{l^{N-1}}.
 \label{eq:rhofinalcp3}
\]

\subsection{Large-$N$ asymptotic form}
\label{sec:largeNcp3}
The large-$N$ analysis can be performed in a similar manner as in the previous sections. Taking the $l^{N-1}$ order 
in \eq{eq:rhofinalcp3} can be represented by a contour integral around the origin, and 
\s[
\left.\left(1+lg\right)^{-2}\, \exp\left(\frac{l}{1+lg}\right)\right|_{l^{N-1}}&=\frac{1}{2 \pi I} \oint_{{\cal C}_0} dl \, l^{-N}
\left(1+lg\right)^{-2}\, \exp\left(\frac{l}{1+lg}\right) \\
&\sim \frac{\tilde g ^N }{I N^N} \oint_{{\cal C}_0} d\tilde l \,\left(1+\tilde l \right)^{-2}\, e^{N f(\tilde l)},
\label{eq:appcontcp3}
\s]
where we have carried out the rescaling of the variables,
\[
l=\frac{\Tilde{l}}{g},\quad g=\frac{\Tilde{g}}{N}\label{eq:scale},
\]
assuming that we focus on the parameter region $g\sim \frac{1}{N}$ as previously, 
and 
\[
f(\tilde l)=-\log \tilde l +\frac{\tilde l}{\tilde g(1+\tilde l)}.
\]
Note that we have ignored some powers of $N$ and $\tilde g$ in the overall factor of \eq{eq:appcontcp3} as
subleadings. 
The saddle points of $f(\tilde l)$ are given by 
\[
      \Tilde{l}^{+}_0=\frac{1-2\Tilde{g}+\sqrt{1-4\Tilde{g}}}{2\Tilde{g}},\quad  \Tilde{l}^{-}_0=\frac{1-2\Tilde{g}-\sqrt{1-4\Tilde{g}}}{2\Tilde{g}}.
\]
We see that the transition point is given by $\tilde g=1/4$, or 
\[
|v|_c=\sqrt{\frac{3 \alpha}{8N}}.
\]

To proceed further, we use the Lefschetz thimble method as in the previous sections. 
 Let us first consider $\Tilde{g}\leq\frac{1}{4}$. In this case the saddle points are on the real axis, and 
 the relevant one we should take can be determined by whether the real part of $f(\tilde l)$ decreases as $\tilde l$ leaves 
 the saddle point in the imaginary direction. In fact we find  
\[
     \left(\frac{\partial^2 f(\tilde l )}{\partial \Tilde{l}^2}\right)_{\tilde l=\tilde l_0^-}\geq 0,
\]
while it is $\leq 0$ for $\tilde l=\tilde l_0^+$. Thus, we shall take the saddle point $\tilde l_0^-$.
Note that, because of $\tilde l_0^->0$, the pole in \eq{eq:appcontcp3} does not contribute unlike in the previous
cases.  Then in the leading order of $N$ we obtain
 \[
 \left. (1+gl)^{-2}\exp\left(\frac{l}{1+gl}\right)\right|_{l^{N-1}}\sim N^{-N} \tilde g^N e^{N f(\tilde l_0^-)}  \,
 \hbox{ for } \tilde g <\frac{1}{4}.
 \label{eq:fgsmall}
 \]
 
 On the other hand, for the case of $\Tilde{g}\geq \frac{1}{4}$, $\tilde l_0^\pm$ are complex and conjugate with 
each other. Therefore we must take both of the solutions, because the result must be real. Thus we obtain
\[
\left. (1+gl)^{-2}\exp\left(\frac{l}{1+gl}\right)\right|_{l^{N-1}}\sim N^{-N} \tilde g^N \left( c_0\, e^{N f(\tilde l_0^+)}+c_0^* \, 
e^{N f(\tilde l_0^-)}\right) \,
 \hbox{ for } \tilde g >\frac{1}{4}
 \label{eq:fglarge}
\]
with a coefficient $c_0$, which is at most in the order of a finite power of $N$.

By putting \eq{eq:fgsmall} and \eq{eq:fglarge} into \eq{eq:rhofinalcp3} and taking into account the factor in the leading order
of $N$, we obtain
\[
\rho_{\rm signed} (|v|)\sim e^{N h},
\label{eq:rhohcp3}
\]
 where 
 \[
h=\log(2)-\frac{\alpha}{|\tilde v|^2} +\re{f(\tilde l_0^-)}.
 \]
In the derivation of \eq{eq:rhohcp3}, we have used 
$c_0 \,e^{Nf(\tilde l_0^+)}+c_0^*\, e^{Nf(\tilde l_0^-)}=2 \re {c_0 e^{I N  \im{f(\tilde l_0^-)}}} e^{N \re{f(\tilde l_0^-)}}$ for
$\tilde g>1/4$, and have ignored the oscillatory factor, since this is at most in the order of a finite power of $N$. 

The edge of the distribution can be determined by $h=0$. Solving numerically, we obtain
\[
|v|_{\rm edge}=0.603501 \sqrt{\frac{\alpha}{N}}.
\]
Note that $h$ is positive real in the region $|v|_{\rm edge}<|v|<|v|_c$. 
By normalizing the eigenvector as $w=v/|v|$, the eigenvector equation \eq{eq:eveqcomplex3} can be rewritten as
\[
C^*_{abc}w_b w_c=z\, w_a^*
\label{eq:cwwws}
\]
with an eigenvalue $z=1/|v|$. Therefore the largest eigenvalue of the complex symmetric order-three random tensor
is asymptotically given by 
\[
z_{\rm largest}=\frac{1}{|v|_{\rm edge}}\sim 1.657 \sqrt{\frac{N}{\alpha}}.
\label{eq:largest}
\] 

The injective norm $|C|_{\rm inj}=\max_{|w|=1} C_{abc}w_a w_b w_c$ of a tensor is an important quantity 
related to the geometric measure of entanglement \cite{shi,barnum,geommeasure,estimate} and  the best rank-one
approximation \cite{SAPM:SAPM192761164,Carroll1970,bestrankone,comon} of a 
tensor. In fact it is straightforward to show
\[
| C |_{\rm inj}=z_{\rm largest}.
\label{eq:cinj}
\]

The injective norm of the complex symmetric order-three random tensor in the large-$N$ asymptotics was numerically 
studied in \cite{estimate}. 
Their random tensor $C$ corresponds to $\alpha=N/2$ of our case, and  their estimated asymptote is $C_0=2.356248$.
This indeed agrees well with $| C |_{\rm inj}=2.34335$ computed from the formula \eq{eq:largest}.

A comment is in order. In the above discussion of comparing our result with the numerical study,  
there are two main assumptions. One is that the edges of the signed and 
the genuine distributions are the same. The other is that the distribution of the smallest eigenvector converges to 
the edge of the distribution in the large-$N$ limit. 
These properties have been proven for the real eigenvalue/vector distribution of the real symmetric random tensor 
\cite{randommat,secondmoment,Kloos:2024hvy}, but not for the other cases including the current case, to the best of 
our knowledge. Though we need to prove these properties to make a final conclusion, we point out
that the present complex case has the following important similarities with the real case:
the signed distribution is monotonic with a constant sign in the region $|v|_{\rm edge} <|v|<|v|_{c}$, 
while it is infinitely oscillatory taking both signs
in the region $|v|>|v|_c$. This is consistent with that only the stable critical points dominate in the region
 $|v|_{\rm edge} <|v|<|v|_{c}$ in the large-$N$ limit, while the others contribute only at $|v|\geq |v|_c$. 
 If so, the signed and the genuine distributions 
are coincident in the region $|v|_{\rm edge} <|v|<|v|_{c}$, and therefore $|v|_{\rm edge}$ is common. 
Moreover, as discussed in \cite{Kloos:2024hvy}, the agreement can naturally be expected, when the eigenvalue/vector
equation can be derived as a critical point equation of a bounded potential. In fact the eigenvalue/vector equation
 \eq{eq:cwwws} can be derived as the critical point equation of a bounded potential,
\[
V= \re{C^*_{abc}w_a w_b w_c} \hbox{ with } |w|=1,
\]
and the largest eigenvalue gives the largest value of $V$. 
In addition we will show a numerical evidence supporting the agreement in 
the right panel of Figure~\ref{fig:mccp3} in Section~\ref{sec:monte}.

\section{Monte Carlo simulations}
\label{sec:monte}

In this section we compare our results with Monte Carlo simulations. We find good agreement between our 
analytical results and the Monte Carlo simulations for all the three cases.
The procedure of the Monte Carlo simulations is basically the same as in our former works 
\cite{Sasakura:2022zwc,Sasakura:2022iqd,Sasakura:2022axo,Sasakura:2023crd,Kloos:2024hvy,Delporte:2024izt,
Sasakura:2024awt}.
We take $\alpha=1/2$ without loss of generality.

We repeat the following random sampling processes.  The number of repetition is denoted by $N_{\rm MC}$. 

\begin{itemize}

\item
Generate the real part (and the imaginary part for the $O(N,\mathbb{C})$ and the $U(N,\mathbb{C})$ cases) 
of $C_{abc}$ by $\sigma/\sqrt{d_{abc}}$, where $\sigma$ 
is randomly generated with the normal distribution of mean value zero and 
standard deviation $1$ (corresponding to $\alpha=1/2$), and $d_{abc}$ is the degeneracy 
defined by\footnote{This degeneracy factor is needed, 
because $C$ has the Gaussian distribution $\propto e^{-\alpha C^2}$ as 
defined in \eq{eq:bareaction}, where $C^2=C_{abc}C_{abc}=\sum_{a\leq b \leq c} d_{abc} C_{abc} C_{abc}$,
because $C$ is a symmetric tensor.} 
\[
d_{abc}=\left\{ 
\begin{array}{ll}
1, & a=b=c, \\
3, & a=b\neq c, \ b=c\neq a, \  c=a\neq b, \\
6, & \hbox{otherwise}.
\end{array}
\right.
\]

\item
Obtain all the complex solutions to the eigenvector equation for a generated $C$. 

\item
Store all the complex solutions. Store also $\det M(v,C)$ for the $U(N,\mathbb{C})$ case.

\end{itemize}

We used a workstation which had a Xeon W2295 (3.0GHz, 18 cores), 128GB DDR4 memory, and Ubuntu 20 as OS.
The eigenvector equations were solved by the NSolve command of Mathematica 14. 

To process the Monte Carlo data obtained by the above sampling processes, 
we divide the parameter space of $v$ into a lattice of bins and count the number of data
which belong to each bin. For the $O(N,\mathbb{R})$ and $O(N,\mathbb{C})$ cases
the parameter space can be taken to be $\vr,\vi,\theta$.
The distribution of the eigenvectors from the Monte Carlo simulations is defined by
\[
\rho_{\rm MC}(\vr,\vi,\theta)=\frac{1}{N_{\rm MC}\Delta_{v_R} \Delta_{v_I} \Delta_\theta} 
\left( {\cal N} (\vr,\vi,\theta)\pm \sqrt{{\cal N} (\vr,\vi,\theta)} \right),
\label{eq:mccp12}
\]
where ${\cal N}(\vr,\vi,\theta)$ denotes the number of the data which belong to 
the bin $(\vr-\Delta_{\vr}/2,\vr+\Delta _{\vr}/2] \times(\vi-\Delta_{\vi}/2,\vi+\Delta_{\vi}/2] 
\times (\theta-\Delta_\theta/2,\theta+\Delta_\theta/2]$, and the last term is an error estimate.
This should be compared with the corresponding analytical expression,
\[
\frac{1}{\Delta_{\vr} \Delta_{\vi} \Delta_{\theta}}
\int_{\vr-\Delta_{\vr}/2}^{\vr+\Delta_{\vr}/2} \int_{\vi-\Delta_{\vi}/2}^{\vi+\Delta_{\vi}/2} 
\int_{\theta-\Delta_\theta/2}^{\theta+\Delta_\theta/2} dx dy dz\, \rho (x,y,z),
\label{eq:intana}
\]
where $ \rho (\vr,\vi,\theta)$ is the analytical result \eq{eq:rhoderexp} or \eq{eq:finalrhocp2} with \eq{eq:rhofinal} 
for the $O(N,\mathbb{R})$ and the $O(N,\mathbb{C})$ cases, respectively. 
The integral \eq{eq:intana} can be replaced by $\rho (\vr,\vi,\theta)$ itself, if it is smooth enough
with respect to the bin size. 

As for the $U(N,\mathbb{C})$ case, the parameter space can be taken to be $|v|$. The difference from 
the other cases is that the sign of $\det M(v,C)$
matters. We have two numbers, ${\cal N}_+(|v|),{\cal N}_{-}(|v|)$,  which count the numbers of data
belonging to the bin $(|v|-\Delta_{|v|}/2,|v|+\Delta_{|v|}/2]$ with $\det M(v,C)>0$ and $\det M(v,C)<0$,
respectively. Then the Monte Carlo signed distribution is defined by 
\[
\rho_{\rm MC\, signed} (|v|)=\frac{1}{N_{\rm MC} \Delta_{|v|}}\left( {\cal N}_+(|v|)- {\cal N}_-(|v|)
\pm \sqrt{{\cal N}_+(|v|)+ {\cal N}_-(|v|)}\right),
\label{eq:mcsigned}
\]
while the Monte Carlo genuine distribution is defined by
\[
\rho_{MC} (|v|)=\frac{1}{N_{\rm MC} \Delta_{|v|}}\left( {\cal N}_+(|v|)+{\cal N}_-(|v|)
\pm \sqrt{{\cal N}_+(|v|)+ {\cal N}_-(|v|)}\right).
\label{eq:mcgenuine}
\]

We can compare \eq{eq:mcsigned} with 
\[
\frac{1}{\Delta_{|v|}} \int_{|v|-\Delta_{|v|}/2}^{|v|+\Delta_{|v|}/2} dx\, \rho_{\rm signed}(x),
\label{eq:anasigned}
\]
where $\rho_{\rm signed}(|v|)$ is given in \eq{eq:rhofinalcp3}.

\begin{figure}
\begin{center}
\includegraphics[width=5cm]{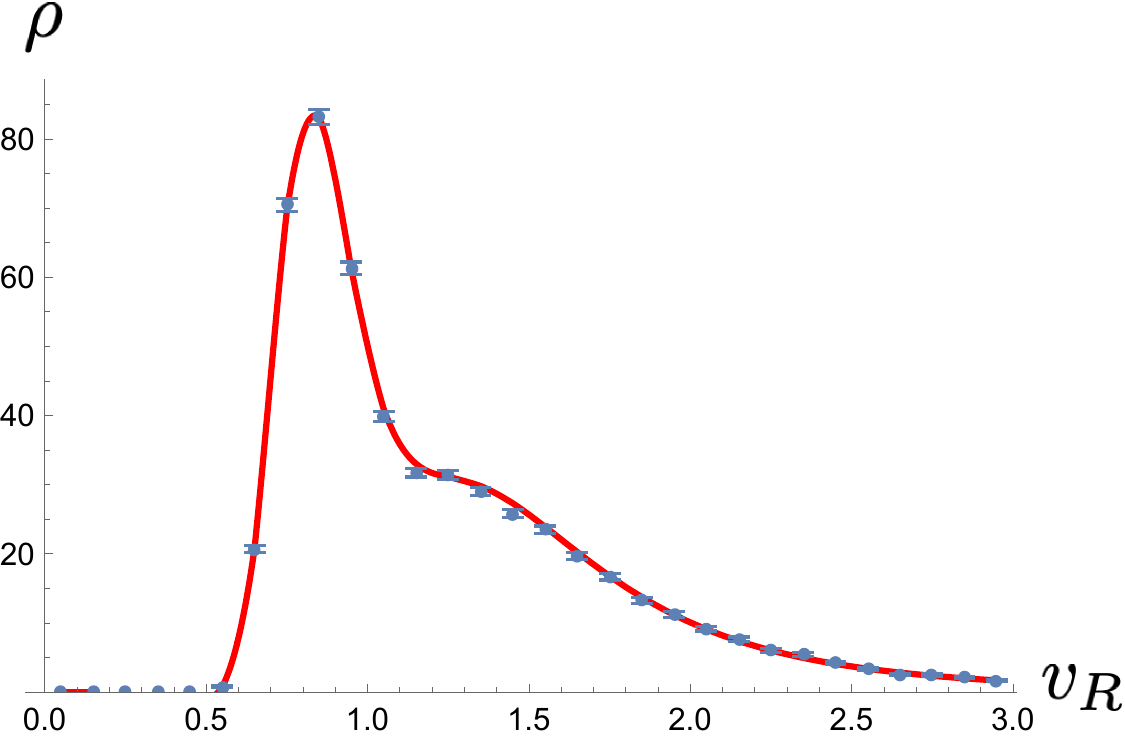}
\hfil
\includegraphics[width=5cm]{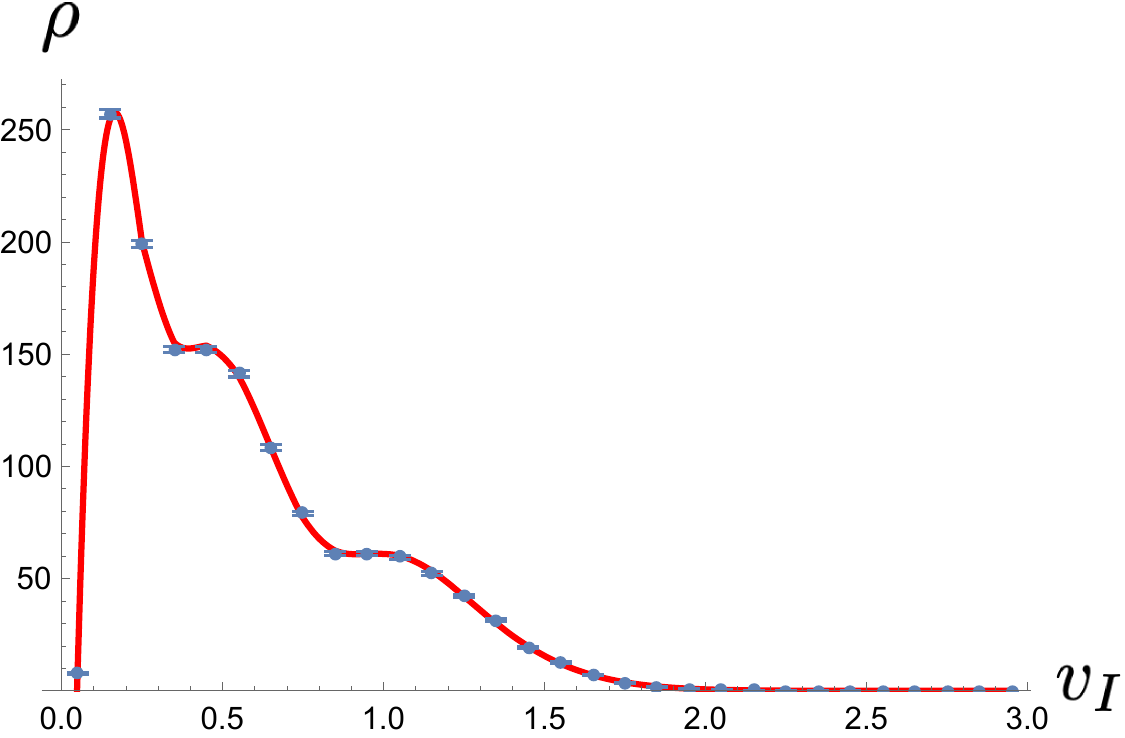}
\hfil
\includegraphics[width=5cm]{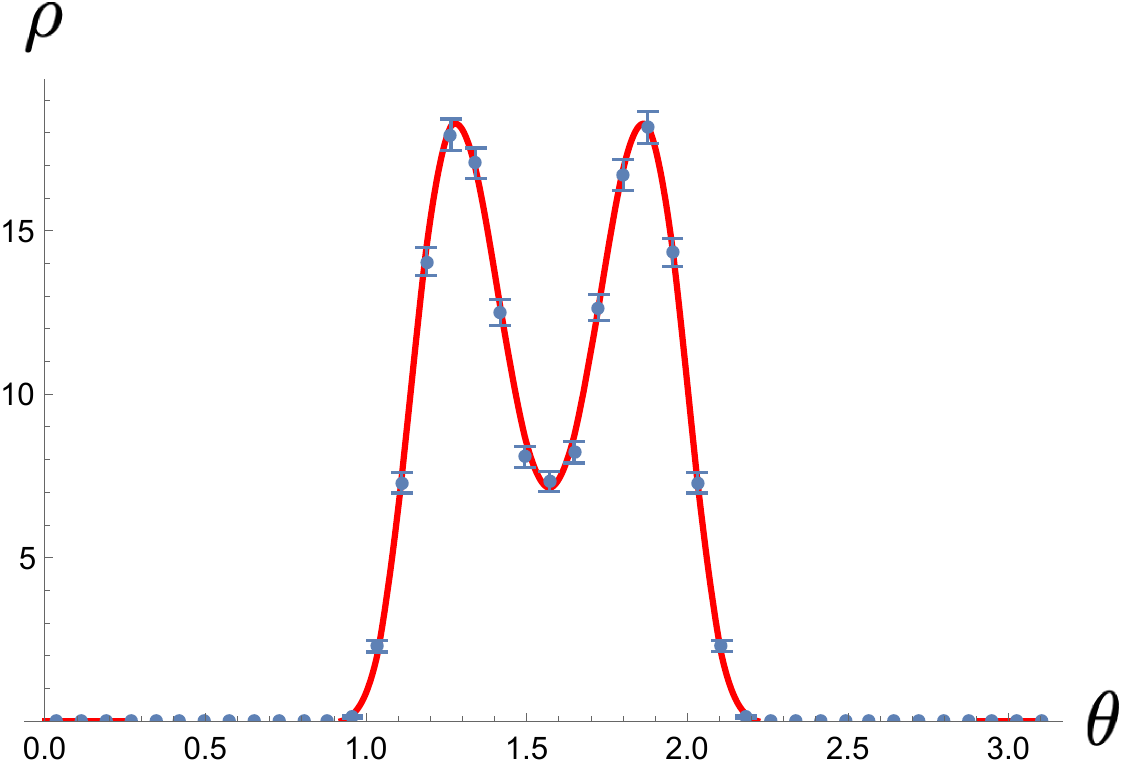}
\caption{The comparison between the Monte Carlo result \eq{eq:mccp12} (dots with error bars)
and the analytical result \eq{eq:rhoderexp} with \eq{eq:rhofinal} (solid line) for the $O(N,\mathbb{R})$ case.
This is for $N=8, N_{MC}=10^5$. The parameters are $\vi=0.95,\theta=39 \pi/82$ (left),
 $\vr=0.95,\theta=39 \pi/82$ (middle), and  $\vr=1.95,\vi=0.55$ (right), with $\Delta_{\vr}=0.1,\Delta_{\vi}=0.1,
 \Delta_\theta=\pi/41$.}
\label{fig:mccp1}
\end{center}
\end{figure} 

\begin{figure}
\begin{center}
\includegraphics[width=5cm]{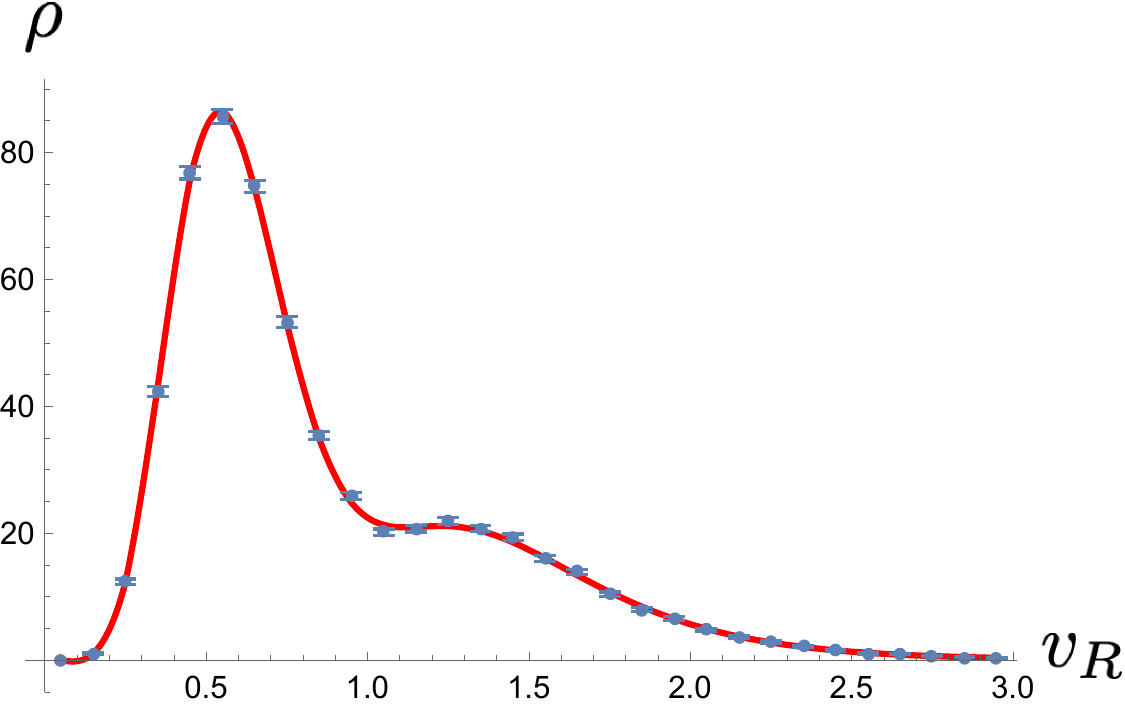}
\hfil
\includegraphics[width=5cm]{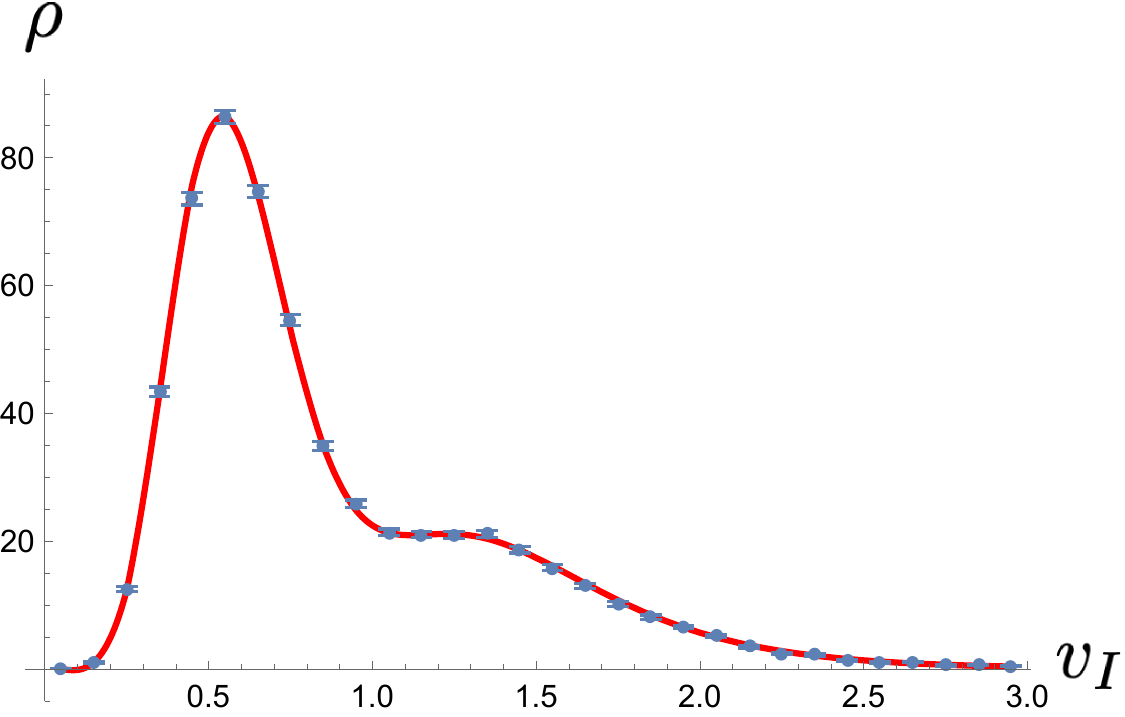}
\hfil
\includegraphics[width=5cm]{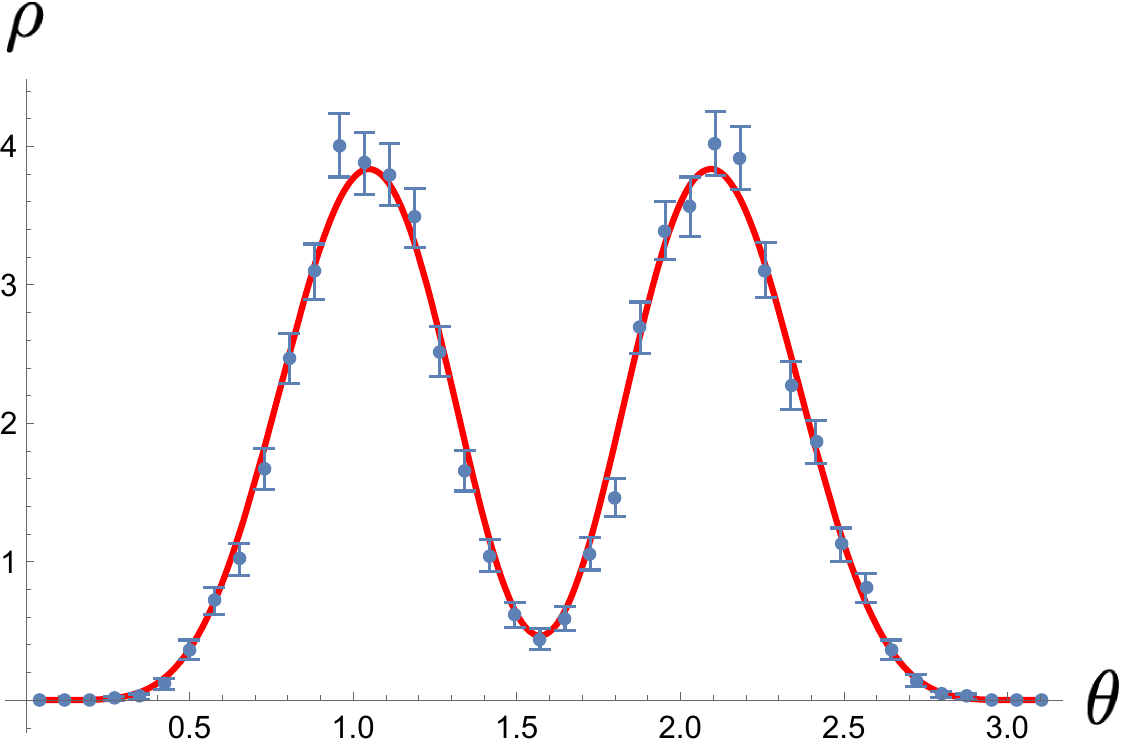}
\caption{The comparison between the Monte Carlo result \eq{eq:mccp12} (dots with error bars) 
and the analytical result \eq{eq:finalrhocp2} with \eq{eq:rhofinal} (solid line) for the $O(N,\mathbb{C})$ case. 
This is for $N=8, N_{MC}=10^5$. The parameters are $\vi=0.95,\theta=39 \pi/82$ (left),
 $\vr=0.95,\theta=39 \pi/82$ (middle), and  $\vr=1.95,\vi=1.95$ (right), with $\Delta_{\vr}=0.1,\Delta_{\vi}=0.1,
 \Delta_\theta=\pi/41$.}
\label{fig:mccp2}
\end{center}
\end{figure}

\begin{figure}
\begin{center}
\includegraphics[width=7cm]{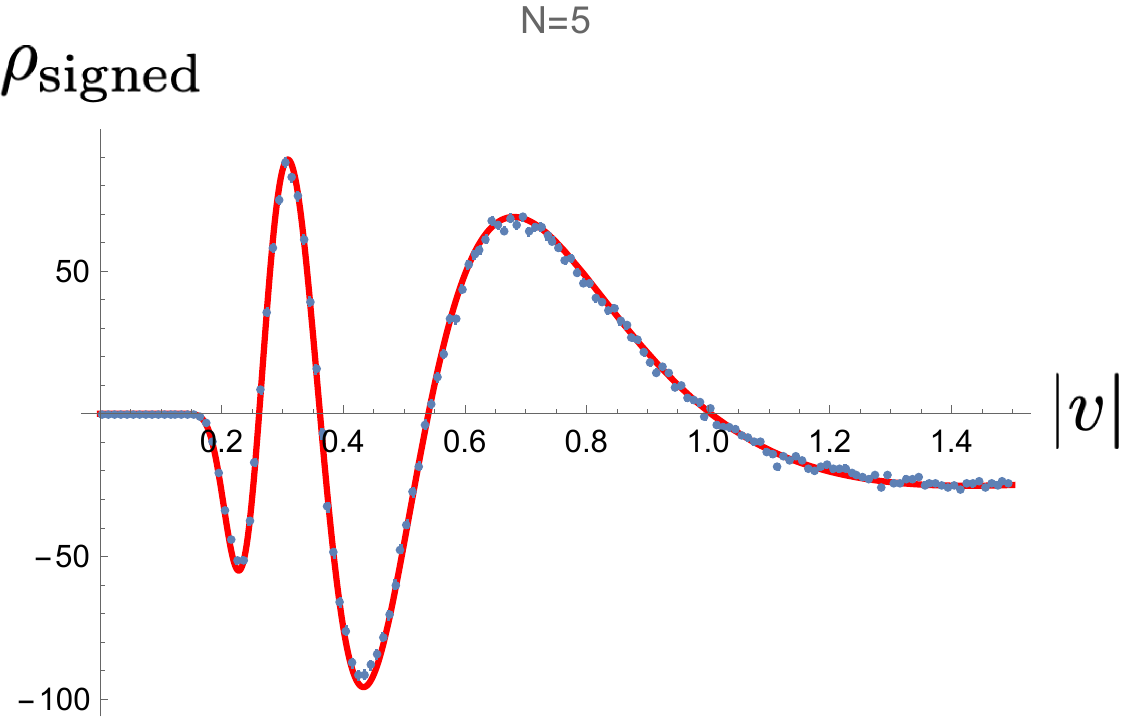}
\hfil
\includegraphics[width=7cm]{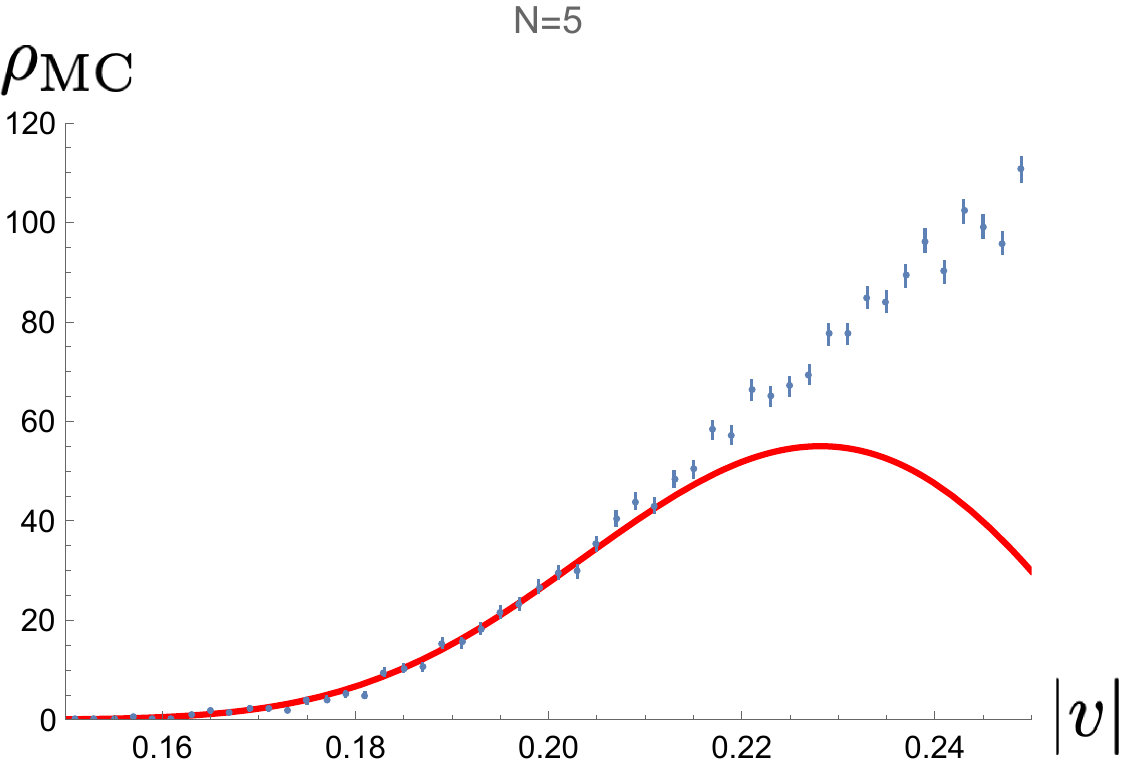}
\caption{Left: The comparison between the Monte Carlo result \eq{eq:mcsigned} (dots with error bars) 
and the analytical result \eq{eq:rhofinalcp3} (solid line) for the $U(N,\mathbb{C})$ case. 
Right: The genuine distribution obtained from the Monte Carlo simulation by \eq{eq:mcgenuine}
is compared with $-\rho_{\rm signed}(|v|)$ in \eq{eq:rhofinalcp3} (solid line). 
They agree well near the end. The parameters are $N=5, N_{\rm MC}=10^4$.}
\label{fig:mccp3}
\end{center}
\end{figure}

In Figure~\ref{fig:mccp1}, we compare the result of the Monte Carlo simulation \eq{eq:mccp12} with the analytical 
result  \eq{eq:rhoderexp} with \eq{eq:rhofinal}  for the $O(N,\mathbb{R})$ case with some examples. 
In Figure~\ref{fig:mccp2}, we do the same for $O(N,\mathbb{C})$.
In the left panel of Figure~\ref{fig:mccp3}, we compare the result of the Monte Carlo simulation \eq{eq:mcsigned} to
\eq{eq:rhofinalcp3} for the $U(N,\mathbb{C})$ case with an example. 
We indeed find very good agreement for all these cases. 

In the right panel of Figure~\ref{fig:mccp3}, we compare the genuine distribution from the Monte Carlo simulation 
\eq{eq:mcgenuine} with $-\rho_{\rm signed}(|v|)$ in \eq{eq:rhofinalcp3}. 
As seen in the plot, they indeed are coincident near the end of the distribution.
This numerically supports the expectation that 
the edge of the signed distribution is in fact the same as that of the genuine distribution in the large-$N$ limit,
as commented in the last paragraph of Section~\ref{sec:largeNcp3}.

Lastly we comment on an issue which we ignored because of its little effect. 
In our Monte Carlo simulations, it is essentially important for the polynomial equation
solver to cover all the solutions to the eigenvector equations. However, in our simulations, 
about 3 percent of them were missed. 
 We could not identify the reason for that, but could estimate the percentage
 from the following facts we noticed.  For the $O(N,\mathbb{R})$ and $O(N,\mathbb{C})$ cases, 
 the total number of non-zero solutions to the eigenvector equation 
 for a randomly generated $C$ agreed with $2^N-1$, 
 which was proven to be the number of eigenvectors in \cite{cart}, but about 3 percent of them were in fact identical. 
 For the $U(N,\mathbb{C})$ case, since the eigenvector equation \eq{eq:eveqcomplex3}
 is invariant under the discrete phase rotation $v\rightarrow e^{2 \pi I/3} v$,
  the solutions should appear in the triplet $(v,e^{2 \pi I/3}v,
 e^{4 \pi I/3}v)$. However, about 3 percent of them did not form this triplet.

 \section{Summary and discussions}
 In this paper we have studied three types of complex eigenvector/value 
 distributions of the complex/real symmetric order-three random tensors,
 where these three cases can be characterized by 
$O(N,\mathbb{R})$, $O(N,\mathbb{C})$, and $U(N,\mathbb{C})$ symmetries, respectively.
In the first two cases the distributions can be represented as partition functions of four-fermi theories,
since the determinants are positive semi-definite.
In the last case, the determinant is not so, 
and we have considered the signed distribution, which can be represented by a four-fermi theory
and is still useful for applications \cite{Kloos:2024hvy,Delporte:2024izt,Sasakura:2024awt}.
We have obtained the exact closed-form expressions of all of these three distributions
by exactly computing the partition functions of the four-fermi theories.
We have taken the large-$N$ limits of these expressions, 
computing the edges and the transition lines/points of the distributions.  
From the edge of the last distribution, we have obtained the injective norm of the complex symmetric order-three 
random tensor, which agrees well with a former numerical study \cite{estimate}.  
 
As already being reported in the previous studies \cite{Kent-Dobias:2020egr,Kent-Dobias2}, 
we see that the edges and the transition lines cross each other
in the holomorphic cases (namely, the $O(N,\mathbb{R})$, $O(N,\mathbb{C})$ cases).
This is different from what happens in the real eigenvalue/vector distribution of the real symmetric random
tensor (referred as RRS below), 
in which there exists the region between the edge and the transition point where locally stable critical points 
dominate \cite{randommat}.  
While this character of RRS can intuitively be understood from the fact that  
the eigenvalue/vector equation of RRS is a critical point equation of a bounded potential,
the holomorphic cases do not indeed have such potentials, as discussed in Section~\ref{sec:absense}.

The third case (namely, the $U(N,\mathbb{C})$ case) 
is not holomorphic and shows the same characteristics as the signed distribution of RRS,
concerning the edge and the transition point \cite{Kloos:2024hvy,Delporte:2024izt,Sasakura:2024awt}.
In the large-$N$ limit, 
the signed distribution of the third case 
is monotonic with a constant sign in the region between the edge and the transition point, 
but is infinitely\footnote{In the limit of large-$N$.} oscillatory taking both signs on the other side of the transition point. 
This is consistent with the following picture proven for RRS \cite{randommat}: In the large-$N$ limit,
locally stable critical pointes dominate in the region between the edge and the transition point,
and the other kinds of critical points contribute only on the other side of the transition point. 
If so, the signed and the genuine distributions agree in the region between the edge and the transition 
point, and in particular they have the common edge of the distributions.
This approves the last statement of the first paragraph: the 
edge of the genuine distribution can instead be computed from the signed distribution. 
We have also numerically shown the coincidence of the signed and the genuine distributions near the edge in 
the right panel of Figure~\ref{fig:mccp3}.

As has been shown in the last case and in the former studies 
\cite{Kloos:2024hvy,Delporte:2024izt,Sasakura:2024awt}, the signed distribution can provide a practical useful method 
for applications, because it has the region of agreement with the genuine distribution at the edge and
is much easer to compute than the genuine distribution. 
However, the agreement of the two has rigorously been proven for RRS only \cite{randommat}, and a rigorous 
generalization is left for future study. 

\vspace{.3cm}
\section*{Acknowledgements}
N.S. is supported in part by JSPS KAKENHI Grant No.19K03825. 

\appendix 
\def\thesection{Appendix  \Alph{section}}

\section{Complex conventions}
\label{app:complex}
The integration measure of a complex variable $x$ is defined by
\[
dx:=dx_R dx_I,
\label{eq:defofcomplexmeasure}
\]
where $x_R=\re{x},x_I=\im{x}$. 
We have 
\s[
\int d\lambda  \, e^{I\lambda f^*+I \lambda^* f}
&=\int_{\mathbb{R}^2} d\lambda_R d\lambda_I  \, e^{2 I (\lambda_R f_R+\lambda_I f_I)}\\
&=\pi^2 \delta(f_R) \delta(f_I),
\label{eq:deltadef}
\s]
where $I$ denotes the imaginary unit, and $f_R=\re{f},f_I=\im{f}$. 

The Gaussian integration over complex values for  a positive-definite $N\times N$ matrix $A$
is given by
\[
\int_{\mathbb{C}^N} dx\, e^{-x^* A x}=\frac{\pi^{N}}{\det A}.
\]

By explicit computation, one finds
\[
\det \left(
\begin{matrix}
\frac{\partial f}{\partial v} &  \frac{\partial f}{\partial v^*} \\
\frac{\partial f^*}{\partial v} & \frac{\partial f^*}{\partial v^*} 
\end{matrix}
\right)
=
\det \left(
\begin{matrix}
\frac{\partial f_R}{\partial v_R} &  \frac{\partial f_R}{\partial v_I} \\
\frac{\partial f_I}{\partial v_R} & \frac{\partial f_I}{\partial v_I} 
\end{matrix}
\right).
\label{eq:detcomp}
\]

\section{Computation of $\det B$}
\label{app:B}
In this appendix we compute the determinant of the matrix $B$ in \eq{eq:defofB}. The determinant of the part of $B$ 
transverse to $v_{i}$ is obviously given by
\[
\det B^\perp=(\det g_2)^{N-2}=(-b^4+a^2 a^{*2})^{N-2}.
\]
As for the parallel part $B^\parallel$, it is easier to compute it by contracting the second index with $v_{ia}$:
\[
B_{ia\,jb} v_{i'a} v_{j'b}=g_{2 ij} g_{i'j'} +2 g_{ij} g_{ji'} g_{ij'},
\]
which is $\tilde B$ defined in \eq{eq:defoftB}.
The determinants are related by $\det \tilde B=\det B^{\parallel} (\det g)^2$. By explicit computation we obtain 
\[
\det \tilde B=9 (b^2-aa^*)^6.
\]
Therefore, 
\[
\det B=\det B^\parallel \det B^\perp=(-1)^N 9  (b^2-aa^*)^{N+2} (b^2+aa^*)^{N-2}.
\]

\section{The real eigenvalue distribution of the real symmetric random tensor}
\label{app:aff}
Various formulas of the complexity of the critical points of the spherical $p$-spin spin-glass model 
are given in \cite{randommat}. The complexity is the same quantity as the distribution of the real eigenvalues 
of the real symmetric random tensor, and we compare it with our result in the $v^I\rightarrow 0$ limit
in Section~\ref{sec:hprofile}. The relation between the variable in \cite{randommat} and ours in the limit 
is given by (See for instance an appendix of \cite{Sasakura:2022axo} for an explicit derivation)
\[
u=-\frac{1}{\tv},
\]
where $u$ is the energy of the spherical $p$-spin spin-glass model.

The $N\rightarrow \infty$ asymptotic formula for the complexity of the critical points with 
index $k$\footnote{The index
$k$ denotes the number of unstable directions of a critical point. The critical points with $k=0$ are the locally stable ones.}
is given by
\[
\Theta_{k,p}(u)=\left\{
\begin{array}{ll}
\frac{1}{2} \log(p-1)-\frac{p-2}{4(p-1)}u^2 -(k+1) I_1(u), &\hbox{if }u\leq -E_{\infty},\\
\frac{1}{2} \log(p-1) -\frac{p-2}{p}, &\hbox{if } u\geq -E_{\infty},
\end{array}
\right.
\]
where $p$ is the order of the random tensor ($p=3$ in our case), $E_{\infty}=2\sqrt{(p-1)/p}$, and
\[
I_1(u)=-\frac{u}{E_\infty^2}\sqrt{u^2-E_\infty^2} -\log(-u+\sqrt{u^2-E_\infty^2})+\log E_\infty.
\]
The asymptotic formula for the complexity of the critical points with no specification of $k$ is given by
\[
\Theta_{p}(u)=\left\{
\begin{array}{ll}
\frac{1}{2} \log(p-1)-\frac{p-2}{4(p-1)}u^2 - I_1(u), &\hbox{if }u\leq -E_{\infty},\\
\frac{1}{2} \log(p-1) -\frac{p-2}{4(p-1)}u^2 , &\hbox{if }  -E_{\infty} \leq u \leq 0,\\
\frac{1}{2} \log (p-1), &\hbox{if }0 \leq u.
\end{array}
\right.
\]

\end{document}